\documentclass[
   aps,
   prl,
   twocolumn,
   reprint,
   superscriptaddress,
   floatfix,
   citeautoscript,
   longbibliography,
   amssymb,
]{revtex4-2}

\usepackage[T1]{fontenc}
\usepackage[utf8]{inputenc}

\usepackage{newtxtext}
\usepackage{newtxmath}

\usepackage{chemformula}
\usepackage{siunitx}
\DeclareSIUnit{\ppma}{ppma}

\usepackage{graphicx}
\graphicspath{{./figures}}

\usepackage[
   unicode,
   colorlinks = true,
   allcolors = blue,
]{hyperref}

\usepackage{cleveref}
\usepackage{microtype}

\usepackage{glossaries}
\glsdisablehyper

\newacronym{1d}{1D}{one-dimensional}
\newacronym{2d}{2D}{two-dimensional}
\newacronym{3d}{3D}{three-dimensional}

\newacronym{ac}{AC}{alternating current}
\newacronym{afm}{AFM}{atomic force microscopy}
\newacronym{alc}{ALC}{avoided level crossing}
\newacronym{api}{API}{application programming interface}
\newacronym{ariel}{ARIEL}{Advanced Rare Isotope Laboratory}
\newacronym{arpes}{ARPES}{angle-resolved photoemission spectroscopy}
\newacronym{atp}{ATP}{adenosine triphosphate}

\newacronym[sort={b-NMR}]{bnmr}{\ensuremath{\beta}-NMR}{\ensuremath{\beta}-detected nuclear magnetic resonance}
\newacronym[sort={b-NQR}]{bnqr}{\ensuremath{\beta}-NQR}{\ensuremath{\beta}-detected nuclear quadrupole resonance}
\newacronym{bca}{BCA}{binary collision approximation}
\newacronym{bcc}{BCC}{body-centred cubic}
\newacronym{bcp}{BCP}{buffered chemical polishing}
\newacronym{bcs}{BCS}{Bardeen-Cooper-Schrieffer}
\newacronym{bpp}{BPP}{Bloembergen-Purcell-Pound}
\newacronym{bsc}{BSC}{\ch{Bi2Se3:Ca}}
\newacronym{btm}{BTM}{\ch{Bi2Te3:Mn}}
\newacronym{bts}{BTS}{\ch{Bi2Te2Se}}

\newacronym{camp}{CAMP}{control and monitor program}
\newacronym{ccd}{CCD}{charge-coupled device}
\newacronym{cdw}{CDW}{charge density wave}
\newacronym{cgs}{CGS}{centimetre-gram-second system of units}
\newacronym{cmms}{CMMS}{Centre for Molecular and Materials Science}
\newacronym{codata}{CODATA}{Committee on Data for Science and Technology}
\newacronym{cpu}{CPU}{central processing unit}
\newacronym{create}{CREATE}{Collaborative Research and Training Experience Program}
\newacronym{cw}{CW}{continuous wave}

\newacronym{daq}{DAQ}{data acquisition}
\newacronym{dc}{DC}{direct current}
\newacronym{dft}{DFT}{density functional theory}
\newacronym{dhva}{dHvA}{de~Haas-van~Alphen}
\newacronym{dos}{DOS}{density of states}
\newacronym{dqt}{DQT}{double-quantum transition}

\newacronym{efg}{EFG}{electric field gradient}
\newacronym{emim-ac}{EMIM-Ac}{1-ethyl-3-methylimidazolium acetate}
\newacronym{emim-dca}{EMIM-DCA}{1-ethyl-3-methylimidazolium dicyanamide}
\newacronym{ep}{EP}{electro-polishing}
\newacronym{epr}{EPR}{electron paramagnetic resonance}
\newacronym{esr}{EPR}{electron spin resonance}
\newacronym{endor}{ENDOR}{electron nuclear double resonance}
\newacronym{epics}{EPICS}{Experimental Physics and Industrial Control System}

\newacronym{fcc}{FCC}{face-centred cubic}
\newacronym{fft}{FFT}{fast Fourier transform}
\newacronym{fom}{FoM}{figure of merit}
\newacronym{fwhm}{FWHM}{full width at half maximum}

\newacronym{gga}{GGA}{generalized gradient approximation}
\newacronym{gl}{GL}{Ginzburg-Landau}

\newacronym{hb}{HB}{hole-burning}
\newacronym{hfqs}{HFQS}{high-field \ensuremath{Q} slope}
\newacronym{hv}{HV}{high-voltage}
\newacronym{hwhm}{HWHM}{half width at half maximum}

\newacronym{iaea}{IAEA}{International Atomic Energy Agency}
\newacronym{icru}{ICRU}{International Commission on Radiation Units and Measurements}
\newacronym{il}{IL}{ionic liquid}
\newacronym{is}{IS}{impedance spectroscopy}
\newacronym{isac}{ISAC}{isotope separator and accelerator}
\newacronym{isol}{ISOL}{isotope separation online}
\newacronym{isosim}{IsoSiM}{Isotopes for Science and Medicine}

\newacronym{lcao}{LCAO}{linear combination of atomic orbitals}
\newacronym{lda}{LDA}{local density approximation}
\newacronym{led}{LED}{light-emitting diode}
\newacronym{leis}{LEIS}{low-energy ion scattering}
\newacronym{lib}{LIB}{lithium-ion battery}
\newacronym{lsat}{LSAT}{\ch{(La,Sr)(Al,Ta)O3}}

\newacronym{mas}{MAS}{magic angle spinning}
\newacronym{mpms}{MPMS}{magnetic property measurement system}
\newacronym{mbe}{MBE}{molecular beam epitaxy}
\newacronym{md}{MD}{molecular dynamics}
\newacronym{midas}{MIDAS}{Maximum Integrated Data Acquisition System}
\newacronym{mit}{MIT}{metal-insulator transition}
\newacronym{mnr}{MNR}{Meyer-Neldel rule}
\newacronym{mqt}{mqt}{multi-quantum transition}
\newacronym{mud}{MUD}{muon data}
\newacronym{ms}{MS}{mass spectrometry}

\newacronym{nbm}{NBM}{neutral beam monitor}
\newacronym{neb}{NEB}{nudged elastic band}
\newacronym{nim}{NIM}{nuclear instrumentation module}
\newacronym{nmr}{NMR}{nuclear magnetic resonance}
\newacronym{no}{NO}{nuclear orientation}
\newacronym{nqr}{NQR}{nuclear quadrupole resonance}
\newacronym{nrc}{NRC}{National Research Council of Canada}
\newacronym{nserc}{NSERC}{Natural Sciences and Engineering Research Council of Canada}

\newacronym{oa}{OA}{optical absorption}

\newacronym{pac}{PAC}{perturbed angular correlation}
\newacronym{pad}{PAD}{perturbed angular distribution}
\newacronym{pas}{PAS}{principle axis system}
\newacronym{pchip}{PCHIP}{piecewise cubic Hermite interpolating polynomial}
\newacronym{pdf}{PDF}{probability density function}
\newacronym{pld}{PLD}{pulsed laser deposition}
\newacronym{pnr}{PNR}{polarized neutron reflectometry}
\newacronym{ppms}{PPMS}{physical property measurement system}
\newacronym{psi}{PSI}{Paul Scherrer Institute}

\newacronym{qens}{QENS}{quasielastic neutron scattering}
\newacronym{ql}{QL}{quintuple layer}
\newacronym{qo}{QO}{quantum oscillations}

\newacronym{rbs}{RBS}{Rutherford backscattering}
\newacronym{rsf}{RSF}{relative sensitivity factor}
\newacronym{rf}{RF}{radio frequency}
\newacronym{rheed}{RHEED}{reflection high-energy electron diffraction}
\newacronym{rib}{RIB}{radioactive ion beam}
\newacronym{rkky}{RKKY}{Ruderman-Kittel-Kasuya-Yosida}
\newacronym{rrr}{RRR}{residual-resistivity ratio}
\newacronym{rtil}{RTIL}{room temperature ionic liquid}

\newacronym{sae}{SAE}{spin-alignment echo}
\newacronym{sans}{SANS}{small angle neutron scattering}
\newacronym{si}{SI}{International System of Units}
\newacronym{sims}{SIMS}{secondary-ion mass spectrometry}
\newacronym{slr}{SLR}{spin-lattice relaxation}
\newacronym{sms}{S\ensuremath{\mu}S}{Swiss Muon Source}
\newacronym[sort={S/N}]{snr}{\textit{S}/\textit{N}}{signal-to-noise ratio}
\newacronym{squid}{SQUID}{superconducting quantum interference device}
\newacronym{srf}{SRF}{superconducting radio frequency}
\newacronym{srim}{SRIM}{Stopping and Range of Ions in Matter}
\newacronym{ssid}{SSID}{solid-state ionic device}
\newacronym{ssr}{SSR}{spin-spin relaxation}
\newacronym{stm}{STM}{scanning tunnelling microscopy}
\newacronym{sts}{STS}{scanning tunnelling spectroscopy}

\newacronym{ti}{TI}{topological insulator}
\newacronym{trim}{TRIM}{Transport and Range of Ions in Matter}
\newacronym{tss}{TSS}{topological surface state}
\newacronym{tmd}{TMD}{transition metal dichalcogenide}

\newacronym{uhv}{UHV}{ultra-high vacuum}

\newacronym{vdw}{vdW}{van der Waals}
\newacronym{vft}{VFT}{Vogel-Fulcher-Tammann}

\newacronym{xrd}{XRD}{x-ray diffraction}
\newacronym{xrr}{XRR}{x-ray reflection}

\newacronym{ybco}{YBCO}{\ch{YBa2Cu3O_{6+x}}}
\newacronym{ysz}{YSZ}{yttria-stabilized zirconia}

\newacronym[sort={muSR}]{musr}{\ensuremath{\mu}SR}{muon spin rotation/relaxation/resonance}
\newacronym{alc-musr}{ALC-\ensuremath{\mu}SR}{avoided level crossing muon spin rotation}
\newacronym{le-musr}{LE-\ensuremath{\mu}SR}{low-energy muon spin spectroscopy}
\newacronym{lf-musr}{LF-\ensuremath{\mu}SR}{longitudinal field muon spin rotation}
\newacronym{rf-musr}{RF-\ensuremath{\mu}SR}{radio frequency muon spin rotation}
\newacronym{tf-musr}{TF-\ensuremath{\mu}SR}{transverse field muon spin rotation}
\newacronym{zf-musr}{ZF-\ensuremath{\mu}SR}{zero field muon spin rotation}

\begin{document}

\title{
	Niobium's intrinsic coherence length and penetration depth revisited using low-energy muon spin spectroscopy
	and secondary-ion mass spectrometry
}


\author{Ryan~M.~L.~McFadden}
\email[E-mail: ]{rmlm@triumf.ca}
\affiliation{TRIUMF, 4004 Wesbrook Mall, Vancouver, BC V6T~2A3, Canada}
\affiliation{Department of Physics and Astronomy, University of Victoria, 3800 Finnerty Road, Victoria, BC V8P~5C2, Canada}

\author{Jonathan~W.~Angle}
\affiliation{Department of Materials Science and Engineering, Virginia Polytechnic Institute and State University, Blacksburg, Virginia 24061, USA}
\affiliation{Pacific Northwest National Laboratory, 902 Battelle Boulevard, Richland, WA 99354, United States of America}

\author{Eric~M.~Lechner}
\affiliation{Thomas Jefferson National Accelerator Facility, 12000 Jefferson Avenue, Newport News, Virginia 23606, United States of America}

\author{Michael~J.~Kelley}
\affiliation{Thomas Jefferson National Accelerator Facility, 12000 Jefferson Avenue, Newport News, Virginia 23606, United States of America}
\affiliation{Nanoscale Characterization and Fabrication Laboratory, Virginia Polytechnic Institute and State University, 1991 Kraft Drive, Blacksburg, Virginia 24061, USA}

\author{Charles~E.~Reece}
\affiliation{Thomas Jefferson National Accelerator Facility, 12000 Jefferson Avenue, Newport News, Virginia 23606, United States of America}

\author{Matthew~A.~Coble}
\affiliation{Pacific Northwest National Laboratory, 902 Battelle Boulevard, Richland, WA 99354, United States of America}

\author{Thomas~Prokscha}
\affiliation{PSI Center for Neutron and Muon Sciences CNM, Paul Scherrer Institute, Forschungsstrasse 111, 5232 Villigen, Switzerland}

\author{Zaher~Salman}
\affiliation{PSI Center for Neutron and Muon Sciences CNM, Paul Scherrer Institute, Forschungsstrasse 111, 5232 Villigen, Switzerland}

\author{Andreas~Suter}
\affiliation{PSI Center for Neutron and Muon Sciences CNM, Paul Scherrer Institute, Forschungsstrasse 111, 5232 Villigen, Switzerland}

\author{Tobias~Junginger}
\email[E-mail: ]{junginger@uvic.ca}
\affiliation{TRIUMF, 4004 Wesbrook Mall, Vancouver, BC V6T~2A3, Canada}
\affiliation{Department of Physics and Astronomy, University of Victoria, 3800 Finnerty Road, Victoria, BC V8P~5C2, Canada}

\date{\today}

\begin{abstract}
We report
direct, simultaneous
measurements of the London penetration depth ($\lambda_L$)
and
\gls{bcs} coherence length ($\xi_{0}$) in oxygen-doped niobium,
with impurity concentrations spanning the ``clean'' to ``dirty'' limits.
Two depth-resolved techniques ---
\gls{le-musr}
and \gls{sims} --- were
used to quantify the element's Meissner screening
profiles,
analyzed within a framework that accounts for nonlocal electrodynamics.
The
analysis indicates intrinsic length scales of
$\lambda_{L} = \qty{29.1 \pm 1.0}{\nano\meter}$
and 
$\xi_{0} = \qty{39.9 \pm 2.5}{\nano\meter}$,
corresponding to a \gls{gl} parameter of
$\kappa = \num{0.70 \pm 0.05}$.
The obtained $\lambda_{L}$ and $\kappa$ values,
accurately quantified at the nanoscale,
are smaller than
those
commonly used in applications and modeling,
and indicate
that clean niobium lies at the boundary between type-I and type-II superconductivity,
supporting the contemporary view that its intrinsic state may be type-I.
\end{abstract}

\maketitle
\glsresetall


A superconductor's fundamental length scales are
closely connected to the material's electronic structure,
with their magnitudes important for understanding the
nature of its superconducting state.
These quantities
---
the London penetration depth $\lambda_{L}$,
denoting the (exponential) decay constant for the attenuation of
magnetic flux density $B$ below a material's surface~\cite{1935-London-PRSLA-149-71},
and
the Pippard/\gls{bcs} coherence length $\xi_{0}$~\cite{1953-Pippard-PRSLA-216-547,1957-Bardeen-PR-108-1175,1971-Halbritter-ZP-243-201},
defining the spatial extent of Cooper pairs (see, e.g.,~\cite{1981-Weisskopf-CP-22-375})
---
govern a superconductor's electrodynamics~\cite{2013-Dressel-ACMP-2013-104379},
which defines salient properties
such as current flow, magnetic screening, and vortex dynamics.
In applied settings,
accurate knowledge of these intrinsic lengths are
essential for the modeling of superconducting electrodynamics
and
the design of devices whose performance depends on them.

With the recent advent of low-energy implanted nuclear-decay spin-probe techniques,
such as
\gls{bnmr}~\cite{2015-MacFarlane-SSNMR-68-1,2022-MacFarlane-ZPC-236-757}
or
\gls{le-musr}~\cite{2004-Bakule-CP-45-203,2004-Morenzoni-JPCM-16-S4583},
it is
possible to \emph{directly} quantify these nanoscale lengths
through measurement of a material's magnetic screening profile.
By implanting spin-active $\beta$-emitters 
(e.g., positive muons $\mu^{+}$)
at \unit{\kilo\electronvolt} energies,
these techniques spatially probe subsurface electromagnetic fields
at depths
on the order of 10s to 100s of nanometers,
comparable to the field-penetration ``layer'' in most superconductors.
To date,
\gls{le-musr} and \gls{bnmr} have been used to quantify Meissner screening in:
the type-I elements
\ch{Pb}~\cite{2004-Suter-PRL-92-087001,2005-Suter-PRB-72-024506,2006-Suter-PB-374-243,2012-Morenzoni-PRB-85-220501}
\ch{Ta}~\cite{2005-Suter-PRB-72-024506,2006-Suter-PB-374-243},
\ch{In}~\cite{2013-Kozhevnikov-PRB-87-104508},
and
\ch{Sn}~\cite{2013-Kozhevnikov-PRB-87-104508};
as well as the
the type-II compounds:
\ch{YBa2Cu3O_{7-$\delta$}}~\cite{2000-Jackson-PRL-84-4958,2004-Khasanov-PRL-92-057602,2010-Kiefl-PRB-81-180502,2013-Saadaoui-PRB-88-180501},
\ch{Ba(Co_{0.074}Fe_{0.926})2As2}~\cite{2012-Ofer-PRB-85-060506},
\ch{Nb3Sn}~\cite{2019-Keckert-SST-32-075004},
$T^{\prime}$-\ch{La_{1.9}Y_{0.1}CuO4}~\cite{2014-Kojima-PRB-89-180508},
and
\ch{NbSe2}~\cite{2009-Hossain-PRB-79-144518}.

Apart from these materials,
perhaps the most well-studied superconductor by implanted spin-probes is
the transition metal \ch{Nb},
owing to its unique properties~\footnote{For example, \ch{Nb} has the highest critical temperature $T_{c} \approx \qty{9.25}{\kelvin}$~\cite{1966-Finnemore-PR-149-231} of the elements and the largest lower critical field $B_{c1} \approx \qty{170}{\milli\tesla}$~\cite{1966-Finnemore-PR-149-231} of all type-II superconductors.}
and
importance for \gls{srf} technology~\cite{2023-Padamsee-SRTA}.
Measurements of the element's Meissner response have been performed on:
``clean''~\cite{2005-Suter-PRB-72-024506} and ``dirty''~\cite{2023-McFadden-JAP-134-163902} thin films;
\gls{srf} cavity cutouts~\cite{2014-Romanenko-APL-104-072601,2024-McFadden-APL-124-086101},
films~\cite{2017-Junginger-SST-30-125013},
and 
surface-treatments~\cite{2023-McFadden-PRA-19-044018};
and as part of a superconducting heterostructure
(see, e.g.,~\cite{2015-DiBernardo-PRX-5-041021,2018-Flokstra-PRL-120-247001,2020-Krieger-PRL-125-026802,2024-Asaduzzaman-SST-37-025002}).
Despite this breadth of study,
questions pertaining to \ch{Nb}'s intrinsic length scales remain.
For example,
\gls{le-musr} measurements on ``clean'' samples~\cite{2005-Suter-PRB-72-024506,2014-Romanenko-APL-104-072601}
find magnetic penetration depths
considerably shorter than the widely cited
$\lambda_{L} \approx \qty{39}{\nano\meter}$~\cite{1965-Maxfield-PR-139-A1515},
which is used extensively in technical applications.
Similarly,
none of these studies~\cite{2005-Suter-PRB-72-024506,2014-Romanenko-APL-104-072601,2017-Junginger-SST-30-125013,2023-McFadden-PRA-19-044018}
have successfully quantified $\xi_{0}$,
due to, in part,
\ch{Nb}'s borderline type-II behavior
(\gls{gl} parameter $\kappa \approx 0.8 > 1/\sqrt{2}$)
that makes its influence on field-screening easily
subdued by modest amounts of impurities.

Thanks to recent advances in understanding the
systematics~\cite{2023-McFadden-PRA-19-044018,2026-McFadden-NIMB-570-165954}
and
subtleties~\cite{2024-McFadden-APL-124-086101,2024-McFadden-AIPA-14-095320}
associated with
screening profile measurements in \ch{Nb},
along with improved methods for quantifying
contaminant species~\cite{2021-Angle-JVSTB-39-024004,2022-Angle-JVSTB-40-024003},
a refined characterization of \ch{Nb}'s superconducting
length scales is now possible.
In this work,
we use a synergistic combination of \gls{sims}
and
\gls{le-musr}
to directly measure the Meissner response in
a series of oxygen-doped \ch{Nb} samples~\cite{2020-Posen-PRA-13-014024,2021-Ito-PTEP-2021-071G01},
whose impurity content has been curated to cover
the ``clean'' and ``dirty'' limits~\cite{2021-Lechner-APL-119-082601,2024-Lechner-JAP-135-133902}.
To extract the length scales from the Meissner profiles,
we use two analysis procedures:
1) a ``staged'' approach,
wherein the subsurface field distribution is fit phenomenologically
and its mean is compared against predictions from an idealized screening model
convolved with the probe's stopping distribution~\cite{2023-McFadden-PRA-19-044018};
and
2) a ``direct'' approach,
wherein the \gls{le-musr} data are fit directly using the field distribution
inferred from the relationship between the probe's stopping distribution
and the Meissner profile model~\cite{2005-Suter-PRB-72-024506}.
In each case,
we treat the character of the element's electrodynamics in both
the local and 
nonlocal limits~\footnote{In a material's Meissner response, nonlocal effects manifest predominantly as: 1) a reduced initial slope in the spatial decay of $B(z)$; and 2) as a negative curvature in $B(z)$, culminating in \emph{sign reversal} when $z \gg \lambda_{L}$. Qualitatively, these can be understood as a consequence of the finite spatial extent of Cooper pairs, causing each ``partner'' to experience different local fields, leading to a diminished screening capacity that is eventually ``overcompensated'' deeper below the surface~\cite{2005-Suter-PRB-72-024506}.},
and
identify \ch{Nb}'s
$\lambda_{L}$ and $\xi_{0}$.
We find the absolute value of $\lambda_{L}$ to be
\qty{\sim 10}{\nano\meter} shorter than the most widely
quoted result~\cite{1965-Maxfield-PR-139-A1515},
but close to predictions from recent electronic structure calculations~\cite{2023-Zarea-FP-11-1269872}
and
other authoritative
measurements~\cite{1999-Benvenuti-PC-316-153,2005-Suter-PRB-72-024506},
whereas our measurement of $\xi_{0}$ is in good agreement with estimates by
others~\footnote{See, for example, the literature average reported in Refs.~\cite{2023-McFadden-PRA-19-044018,2023-McFadden-JAP-134-163902}}.
Implications of the updated values are discussed,
focussing on \ch{Nb}'s intrinsic superconducting type classification
and its use in \gls{srf} cavities.


\begin{table}
	\caption{
		\label{tab:samples}
		Summary of the near-surface impurity content in the oxygen-doped \ch{Nb}
		samples derived from \gls{sims}.
		Here,
		$[i]$ denotes the concentration of each major contaminant species
		($i = \ch{C}, \ch{N}, \ch{O}$),
		and
		$\ell$ is the corresponding
		electron mean-free-path calculated from \Cref{eq:mfp}~\cite{1976-Schulze-ZM-67-737,1968-Goodman-JPF-29-240}.
		The range of impurity content roughly spans \ch{Nb}'s ``clean''
		and ``dirty'' limits.
	}
	\begin{tabular*}{\columnwidth}{l @{\extracolsep{\fill}} S S S S}
\botrule
{Sample} & {$[\ch{C}]$ (ppma)} & {$[\ch{N}]$ (ppma)} & {$[\ch{O}]$ (ppma)} & {$\ell$ (nm)} \\
\hline
Nb-SR12 & 123 \pm 17 & 54 \pm 9 & 2540 \pm 190 & 30.2 \pm 2.9 \\
Nb-SR13 & 113 \pm 17 & 51 \pm 7 & 50 \pm 5 & 380 \pm 40 \\
Nb-SR14 & 115 \pm 11 & 45.4 \pm 2.9 & 2310 \pm 120 & 33.3 \pm 2.6 \\
Nb-SR15 & 154.1 \pm 3.1 & 41.3 \pm 3.3 & 5840 \pm 350 & 13.6 \pm 1.2 \\
Nb-SR16 & 127 \pm 18 & 46 \pm 4 & 4200 \pm 210 & 18.8 \pm 1.5 \\
Nb-SR18 & 101 \pm 6 & 46.2 \pm 3.2 & 1041 \pm 34 & 69 \pm 5 \\

\botrule
\end{tabular*}
\end{table}

\ch{Nb} samples were cut from a single high
\gls{rrr} (i.e., \num{> 300}) sheet.
The surface of each sample
(\qty{\sim 2.5}{\centi\meter} diameter discs)
was prepared using a combination of chemical etching,
mechanical polishing,
and vacuum annealing to produce a defect-free surface in accord with
high-performance \gls{srf} cavity fabrication standards~\cite{2023-Lechner-PRAB-26-103101}.
Following these standard preparation steps,
select samples were anodized to control the thickness of \ch{Nb}'s native surface oxide layer~\cite{1987-Halbritter-APA-43-1}.
The samples then received a \qtyrange{300}{350}{\celsius} vacuum heat treatment
(often referred to as a ``mid-$T$'' bake for \gls{srf} cavities~\cite{2020-Posen-PRA-13-014024})
to modify the near-surface oxygen impurity content via a dissolution process~\cite{1990-King-TSF-192-351},
followed by light \gls{ep} to remove residual contaminants
and
ensure uniform impurity concentration throughout the Meissner screening region.
Quantitative \gls{sims} measurements~\cite{2021-Angle-JVSTB-39-024004,2022-Angle-JVSTB-40-024003}
on identically prepared ``companion'' samples
were used to identify impurity species and their concentrations.
Prior to measurement, each ``companion'' was dosed with a low-abundance isotope
for each elemental impurity (e.g., \ch{^{18}O}),
whose profile was used as an \emph{in situ} standard
(i.e., to mitigate sample-to-sample \gls{rsf} variations).
Each sample's electron mean-free-path $\ell$ was calculated
using~\cite{1968-Goodman-JPF-29-240,1976-Schulze-ZM-67-737}:
\begin{equation}
	\label{eq:mfp}
	\ell \approx \frac{ \qty{3.7e-16}{\ohm\meter\squared} }{ \sum_{i} a_{i} \cdot [i] } ,
\end{equation}
where $[i]$ is the concentration of impurity element $i$
and $a_{i}$ is an empirical proportionality constant~\cite{1976-Schulze-ZM-67-737}.  
The results are summarized in \Cref{tab:samples}
and further details can be found in the
Supporting Material~\cite{sm}.

\begin{figure}
	\centering
	\includegraphics[width=1.0\columnwidth]{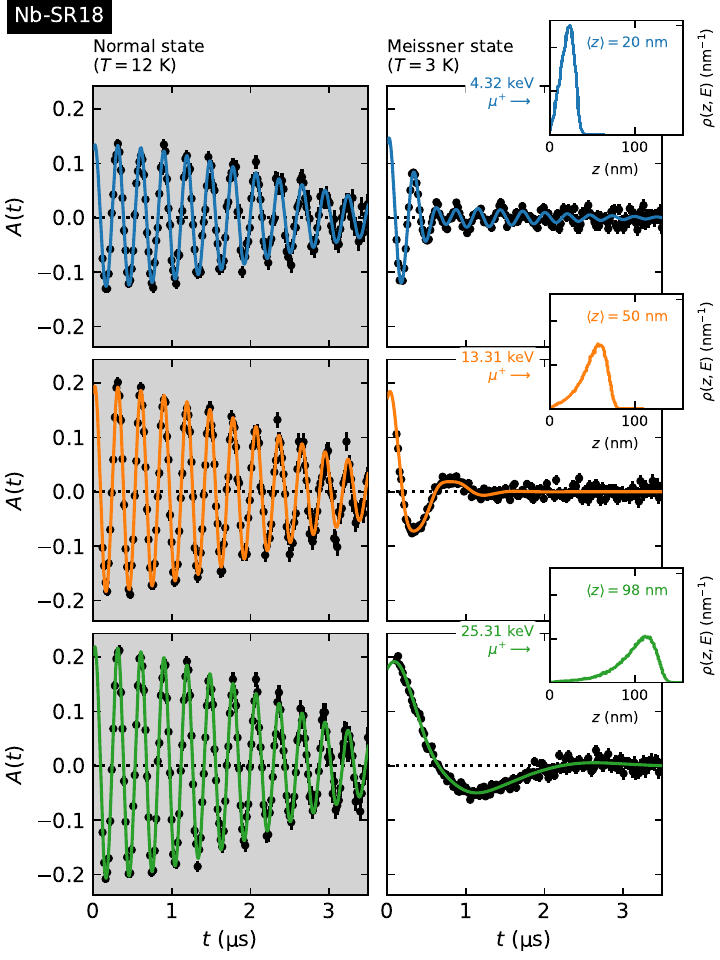}
	\caption{
		\label{fig:asymmetry}
		Typical \gls{le-musr} data in \ch{Nb} (sample \ch{Nb}-SR18).
		In the normal state,
		the $\mu^{+}$ asymmetry $A(t)$ is weakly damped
		with a spin-precession rate that is independent of
		implantation energy $E$.
		By contrast,
		the damping of $A(t)$ in the Meissner state
		is strong, increasing with increasing $E$,
		which is accompanied by a decrease in the
		rate of spin-precession.
		The solid colored lines denote fits to a model approximating the field distribution
		as a sum of Gaussians
		(described in the text and Supporting Material~\cite{sm}).
		The $\mu^{+}$ implantation profile $\rho(z, E)$
		and
		mean stopping depth $\langle z \rangle$,
		simulated using the
		\texttt{TRIM.SP} Monte Carlo code~\cite{1991-Eckstein-SSMS-10,trimsp},
		are shown in the inset for each $E$.
	}
\end{figure}

\Gls{le-musr} measurements were performed at the \gls{sms}
(located at the \gls{psi} in Villigen, Switzerland)
on the
$\mu E4$ beamline~\cite{2008-Prokscha-NIMA-595-317}.
A \qty{\sim 100}{\percent} spin-polarized low-energy
(\qty{\sim 15}{\kilo\electronvolt})
$\mu^{+}$ beam
was generated by energy-moderating a ``surface'' $\mu^{+}$ source
using a condensed cryogenic gas film~\cite{1994-Morenzoni-PRL-72-2793,2001-Prokscha-ASS-172-235}.
The beam was delivered electrostatically to a dedicated spectrometer~\cite{2000-Morenzoni-PB-289-653,2008-Prokscha-NIMA-595-317,2012-Salman-PP-30-55}
via an \gls{uhv} beamline.
An electrically isolated sample holder connected to a bipolar \gls{hv} power supply
was used to control the beam's implantation energy,
allowing the $\mu^{+}$ stopping depth to be controlled
between \qtyrange{\sim 10}{\sim 150}{\nano\meter}~\cite{2002-Morenzoni-NIMB-192-245,2023-McFadden-PRA-19-044018}.
This stopping process was simulated using the Monte Carlo code
\texttt{TRIM.SP}~\cite{1991-Eckstein-SSMS-10,trimsp}
with typical implantation profiles shown in \Cref{fig:asymmetry}
and
in the Supporting Material~\cite{sm}.
Note that these simulations make use of refinements to the parameterization of 
\ch{Nb}'s electronic stopping cross section~\cite{2023-McFadden-PRA-19-044018,2026-McFadden-NIMB-570-165954},
which were recently confirmed to give the most accurate account
of $\mu^{+}$'s range~\cite{2026-McFadden-NIMB-570-165954}.

The Meissner screening profile in each \ch{Nb} sample
was measured by \gls{le-musr} using a transverse-field geometry,
wherein an external field $B_{\mathrm{applied}} \approx \qty{25}{\milli\tesla}$
was applied perpendicular to the initial direction of $\mu^{+}$ spin-polarization
and parallel to the sample surface.
This configuration is highly sensitive to field inhomogeneities,
allowing for the local field distribution $p(B)$,
which is directly related to \ch{Nb}'s Meissner screening profile $B(z)$,
to be determined.
In this setup,
the $\mu^{+}$ spin-polarization $P_{\mu}(t)$ follows (see, e.g.,~\cite{2024-Amato-IMSS}):
\begin{equation}
	\label{eq:polarization}
	P_{\mu}(t) = \int_{0}^{\infty} p(B) \cos \left ( \omega_{\mu} t + \phi_{\pm} \right ) \, \mathrm{d} B , 
\end{equation}
where $t$ is the time after implantation,
$\phi_{\pm}$ is a (detector-dependent) phase factor,
and
$\omega_{\mu} = \gamma_{\mu} B$ is the probe's Larmor frequency,
with
$\gamma_{\mu} / (2 \pi) = \qty{135.539}{\mega\hertz\per\tesla}$
denoting its gyromagnetic ratio~\cite{2021-Tiesinga-RMP-93-025010}.
This process is monitored via the anisotropic emissions from $\mu^{+}$ $\beta$-decay
(mean lifetime $\tau_{\mu} = \qty{2.197}{\micro\second}$~\cite{2022-Workman-PTEP-2022-083C01-short}),
wherein the direction of an emitted $\beta$-ray is correlated with
the spin direction at the moment of decay.
Specifically,
one measures the \emph{asymmetry} $A(t) \equiv A_{0} P_{\mu}(t)$ between a set of opposing radiation counters,
where $A_{0}$ is a proportionality constant
(see the Supporting Material~\cite{sm}).
All Meissner state measurements were performed at \qty{3}{\kelvin} following a zero-field-cooling procedure.
Typical transverse-field \gls{le-musr} data in \ch{Nb}
are shown in \Cref{fig:asymmetry}.

\begin{figure}
	\centering
	\includegraphics[width=1.0\columnwidth]{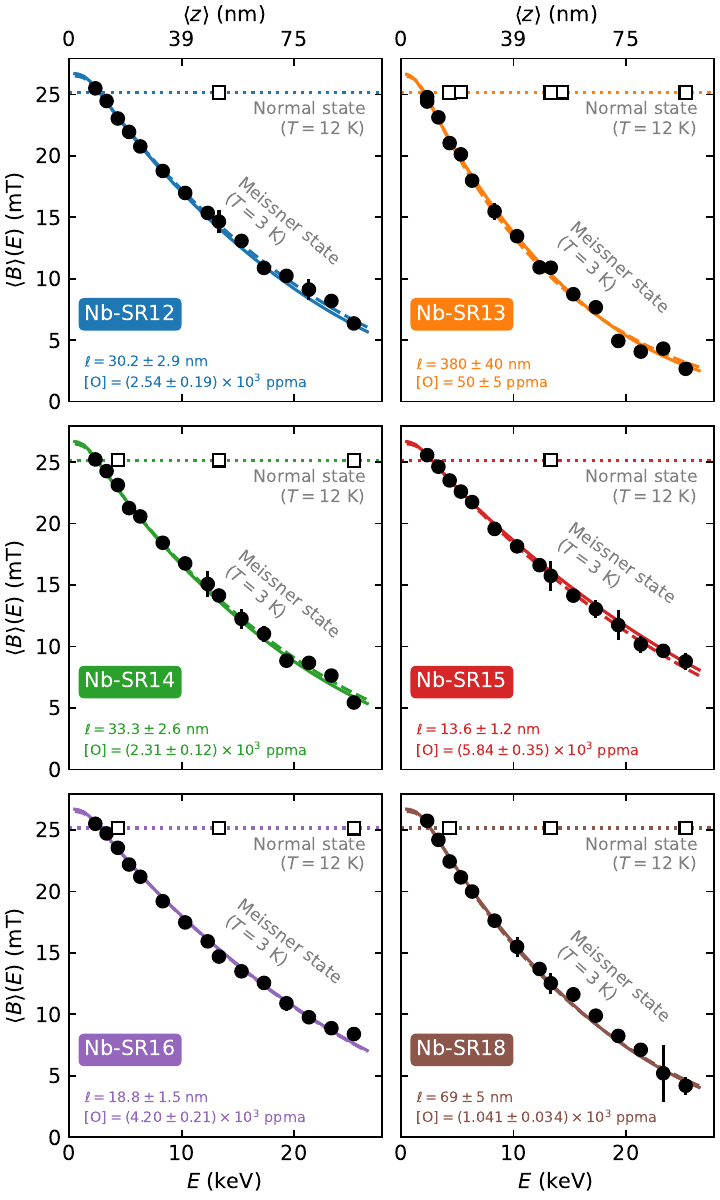}
	\caption{
		\label{fig:profiles}
		Meissner screening profiles in each \ch{Nb} sample,
		derived from a ``staged'' analysis of the \gls{le-musr} data.
		Here,
		the mean magnetic field $\langle B \rangle$ is plotted
		as a function of $\mu^{+}$ implantation energy $E$
		(with the corresponding mean stopping depth $\langle z \rangle$ also indicated)
		in both the normal and Meissner states.
		Differences in screening capacity are visually evident,
		with each sample's oxygen impurity concentration $[\ch{O}]$
		and electron mean-free-path $\ell$
		(derived from \gls{sims})
		indicated in each panel. 
		The solid and dotted colored lines represent a global fit of the ``staged'' data
		to the nonlocal model for the screening profile $B(z)$
		convolved with the $\mu^{+}$ stopping distribution $\rho(z,E)$
		(described in the text).
		For comparison,
		the dashed colored lines show similar fits assuming $B(z)$ is governed
		by local electrodynamics.
	}
\end{figure}

Strong changes in $P_{\mu}(t)$,
and hence $p(B)$,
are evident upon transitioning from the normal to Meissner state.
In the normal state,
long-lived coherent spin-precession is observed,
consistent with a narrow Gaussian $p(B)$.
Conversely,
the strong damping and beating signal in the Meissner state suggests a
much broader $p(B)$,
which is expected to be asymmetric
(i.e., skewed to lower fields).
For \ch{Nb},
this behavior can be approximated by either a skewed Gaussian
distribution~\cite{2023-McFadden-PRA-19-044018,2024-Asaduzzaman-SST-37-025002}
or a sum of Gaussian distributions,
with fits of the \gls{le-musr} data to the latter model
(performed using \texttt{musrfit}~\cite{2012-Suter-PP-30-69})
shown \Cref{fig:asymmetry}.
Further analysis details,
along with similar fits using the ``direct'' approach,
can be found in the Supporting Material~\cite{sm}.


To extract the screening lengths from the ``staged'' analysis,
for each of the \gls{le-musr} measurements
we identify the mean magnetic field $\langle B \rangle$ of the measured $p(B)$
and
plot its
dependence on implantation energy $E$
(see \Cref{fig:profiles}).
As expected,
$\langle B \rangle$ is $E$-independent
(i.e., depth-independent)
in the normal state,
but decays substantially in the Meissner state with increasing $E$.
The form of this decay is related to the Meissner profile $B(z)$ through:
\begin{equation}
	\label{eq:average-field}
	\langle B \rangle (E) = \int_{0}^{\infty} B(z) \rho(z, E) \, \mathrm{d}z ,
\end{equation}
where $\rho(z, E)$ is the $E$-dependent $\mu^{+}$ stopping profile
(see \Cref{fig:asymmetry})~\footnote{As the $\rho(z, E)$s are obtained from Monte Carlo simulations performed at discrete $E$, here we adopt an empirical ``interpolation'' scheme (see~\cite{2023-McFadden-PRA-19-044018,2024-Asaduzzaman-SST-37-025002,2024-McFadden-APL-124-086101} and the Supporting Material~\cite{sm}) to facilitate evaluation of \Cref{eq:average-field} at \emph{arbitrary} $E$.}.
For a semi-infinite slab geometry with specular surface reflection~\footnote{The assumption of mirror-like reflection of electrons at a superconductor's surface can be contrasted with diffusive scattering. In reality, most materials likely exhibit behavior between these two limits. Pragmatically, either choice is permissible as both produce nearly identical screening lengths~\cite{1996-Tinkham-IS-2}.},
the screening profile can be written as~\cite{1996-Tinkham-IS-2}:
\begin{equation}
	\label{eq:nonlocal-screening}
	B(z) = \tilde{B}_{0} \times \begin{cases}
		1 & z < d, \\
		\displaystyle \frac{2}{\pi} \int_{0}^{\infty} \frac{q \sin \left ( q \left [z - d \right ] \right ) }{q^{2} + K(q)} \, \mathrm{d}q & z \geq d , \\
	\end{cases} 
\end{equation}
where
$z$ is the depth below the surface,
$d$ is an empirical parameter accounting for a finite non-superconducting
surface ``dead layer''~\footnote{While the ``dead layer'' $d$ is a sample-dependent quantity, non-zero values are common for \ch{Nb}~\cite{2005-Suter-PRB-72-024506,2014-Romanenko-APL-104-072601,2017-Junginger-SST-30-125013,2023-McFadden-PRA-19-044018,2023-McFadden-JAP-134-163902}, likely due to its natural surface oxidation~\cite{1987-Halbritter-APA-43-1}.},
and
$\tilde{B_{0}}$ accounts for any geometric enhancement to
$B_{\mathrm{applied}}$:
\begin{equation}
	\label{eq:field-enhancement}
	\tilde{B}_{0} = B_{\mathrm{applied}} \times \begin{cases}
		1 & T \geq T_{c} , \\
		\left (1 - \tilde{N} \right )^{-1} & T \ll T_{c} , \\
	\end{cases}
\end{equation}
with $\tilde{N}$ denoting the sample's effective demagnetization
factor~\footnote{Even though $B_{\mathrm{applied}}$ is parallel the sample surface in the measurements, the finite thickness of our samples (i.e., a few millimters) is enough to ensure that $\tilde{N} > 0$ (cf.\ thin film samples where $\tilde{N} \rightarrow 0$~\cite{2005-Suter-PRB-72-024506}).}.
The remaining term $K(q)$ in \Cref{eq:nonlocal-screening}
is the Fourier transformed
(i.e., wavevector $q$ dependent) 
integrand kernel for the nonlocal relationship
between the current density $\mathbf{j}$
and
magnetic vector potential $\mathbf{A}$
(see, e.g.,~\cite{2005-Suter-PRB-72-024506}).
Following a modern version~\cite{2005-Suter-PRB-72-024506,1996-Tinkham-IS-2} of Pippard's model~\cite{1953-Pippard-PRSLA-216-547},
$K(q)$ can be defined analytically as~\footnote{Non-analytic expressions equivalent to \Cref{eq:nonlocal-kernel} from Pippard theory~\cite{1953-Pippard-PRSLA-216-547} can be derived from \gls{bcs} theory~\cite{1957-Bardeen-PR-108-1175,1971-Halbritter-ZP-243-201}. Importantly, quantitative differences between the two models are negligible~\cite{2005-Suter-PRB-72-024506}.}:
\begin{equation}
	\label{eq:nonlocal-kernel}
	K(q) = \frac{ \xi(T) }{ \lambda(T)^{2} \xi(0) } \left \{ \frac{3}{2} \frac{ g(x) }{ x^{3} } \right \} ,
\end{equation}
where
$x \equiv q \xi(T)$,
$g(x) \equiv ( 1 + x^{2}) \arctan ( x ) - x$,
$\lambda(T)$ is the magnetic penetration depth~\footnote{Note that the ``two-fluid'' temperature dependence in \Cref{eq:penetration-depth} is known to deviate from \gls{bcs} theory~\cite{1957-Bardeen-PR-108-1175} when $T \ll T_{c}$; however, it is known to accurately reproduce \ch{Nb}'s $\lambda(T)$ (see, e.g.,~\citenum{2005-Suter-PRB-72-024506}).}:
\begin{equation}
	\label{eq:penetration-depth}
	\lambda (T) \approx \frac{ \lambda_{L} }{ \sqrt {1 - (T / T_{c})^{4} } } ,
\end{equation}
and
$\xi(T)$ is the coherence length~\footnote{Note that the factor of $(\pi / 2)$ included in \Cref{eq:coherence-length} ensures equivalent length scales with \gls{bcs} theory~\cite{1971-Halbritter-ZP-243-201}.}:
\begin{equation}
	\label{eq:coherence-length}
	\frac{1}{\xi(T)} = \frac{J(0, T)}{ \left ( \pi / 2 \right ) \xi_{0}} + \frac{1}{\ell} .
\end{equation}
The weak temperature dependence of \Cref{eq:coherence-length} is dictated by the
\gls{bcs} ``range'' function
(i.e., the real-space kernal obtained via Fourier transform of $K(q)$)~\cite{1957-Bardeen-PR-108-1175,1996-Tinkham-IS-2}:
\begin{equation}
	\label{eq:bcs-range}
	J(0, T) = \left \{ \frac{ \lambda(T) }{ \lambda(0) } \right \}^{2} \frac{ \Delta(T) }{ \Delta(0) } \tanh \left \{ \frac{ \Delta(T) }{ 2 k_{B} T } \right \} ,
\end{equation}
where $\Delta(T) \approx \Delta_{0} \sqrt{ \cos [ 0.5 \pi  (T / T_{c})^{2} ]  }$ is the superconducting energy gap~\cite{1966-Sheahen-PR-149-368}
and $\Delta_{0}$ is its value at \qty{0}{\kelvin}.

\begin{figure}
	\centering
	\includegraphics[width=1.0\columnwidth]{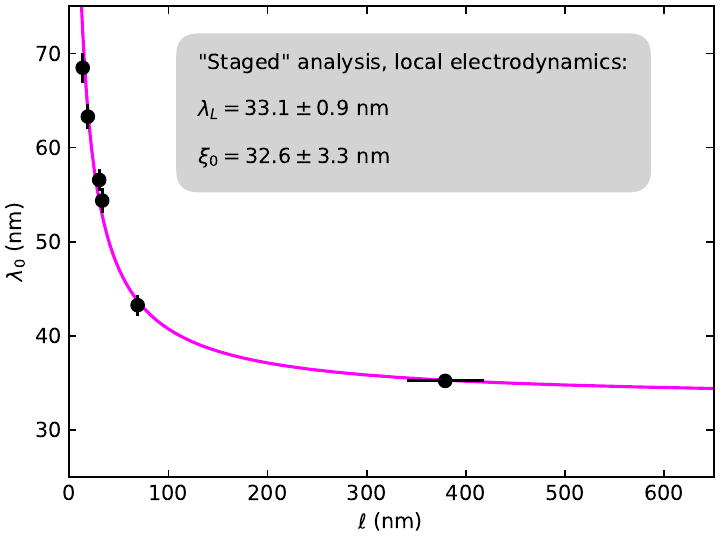}
	\caption[Determination of $\lambda_{L}$ and $\xi_{0}$ for the ``staged'' analysis approach in the local limit.]{
		\label{fig:lambda-vs-mfp-staged}
		Dependence of the (effective) magnetic penetration depth at \qty{0}{\kelvin},
		determined the ``staged'' analysis approach in the local limit,
		on the electron mean-free-path $\ell$.
		The solid coloured line denotes a best fit to \Cref{eq:lambda-impurity},
		with corresponding values for the London penetration depth $\lambda_{L}$ and
		the Pippard/\gls{bcs} coherence length indicated in the plot.
	}
\end{figure}

Using \Cref{eq:average-field,eq:nonlocal-screening,eq:field-enhancement,eq:nonlocal-kernel,eq:penetration-depth,eq:coherence-length,eq:bcs-range},
along with the empirical parameterization of $\rho(z,E)$
(see the Supporting Material~\cite{sm}),
we fit the $\langle B \rangle$ vs. $E$ data in each sample
using a simultaneous (i.e., global) minimization routine
with shared parameters
$\lambda_{L}$,
$\xi_{0}$,
$B_{\mathrm{applied}}$,
and
$\tilde{N}$~\footnote{From fitting, we find that the effective demagnetization factor $\tilde{N}$ ranges from \numrange{0.03}{0.06}, with its exact value depending on the sample and analysis model. The ``staged'' analysis generally finds $\tilde{N} \gtrsim 0.05$, while $\tilde{N} \lesssim 0.04$ is found in the ``direct'' approach. Exact values are given in Tables~S4-S7 in the Supporting Material~\cite{sm}.},
as well as common fixed values
$T_{c} = \qty{9.25}{\kelvin}$~\cite{1966-Finnemore-PR-149-231}
and
$\Delta(0) = \qty{1.53}{\milli\electronvolt}$~\cite{1962-Townsend-PR-128-591}.
The fit results are shown in \Cref{fig:profiles},
in good agreement with the data.
Similar fits assuming local electrodynamics
(i.e., $K(q) \approx \{ \lambda_{0} / \sqrt{1 - (T / T_{c})^{4} } \}^{-2}$, where $\lambda_{0}$ is the effective penetration depth at \qty{0}{\kelvin} --- see the Supporting Material~\cite{sm})
are also shown in \Cref{fig:profiles},
deviating only slightly from the nonlocal result.
The variation of the extracted $\lambda_{0}$ with $\ell$ is shown in
\Cref{fig:lambda-vs-mfp-staged},
with the trend following~\cite{1959-Miller-PR-113-1209,1971-Halbritter-ZP-243-201}:
\begin{equation}
	\label{eq:lambda-impurity}
	\lambda_{0} \approx \lambda_{L} \sqrt{ 1 + \frac{ \pi \xi_{0} }{ 2 \ell } } .
\end{equation}
For comparison,
we applied both the local and nonlocal models using the ``direct'' approach
(see the Supporting Material~\cite{sm})
with the $\lambda_{L}$ and $\xi_{0}$ extracted from each methodology collated
in \Cref{tab:results}
(tables summarizing all fit results from the above analyses are given in the Supporting Material~\cite{sm}).
Though small differences in each length are evident,
their
magnitude is insensitive to the analysis model chosen,
affirming the consistency of our approach.
We shall examine both quantities in detail below.

\begin{table}
	\centering
	\caption{
		\label{tab:results}
		Summary of estimates
		for \ch{Nb}'s intrinsic
		London penetration depth $\lambda_{L}$
		and
		\gls{bcs}/Pippard coherence length $\xi_{0}$,
		obtained from different analysis approaches
		(described in the text and Supporting Material~\cite{sm}).
		Derived values for the element's
		\gls{gl} parameter $\kappa$
		are also given.
		Weighted averages (w.a.) of the $\lambda_{L}$ and $\xi_{0}$ values in the local and nonlocal limits are also included,
		along with the corresponding $\kappa$.
		We take the values derived from the nonlocal (w.a.) as the best estimates for \ch{Nb}.
	}
\begin{tabular*}{\columnwidth}{l @{\extracolsep{\fill}} l S S S}
\botrule
{Method} & {} & {$\lambda_{L}$ (\unit{\nano\meter})} & {$\xi_{0}$ (\unit{\nano\meter})} & {$\kappa$} \\
\hline
Staged & Local & 33.1 \pm 0.9 & 32.6 \pm 3.3 &  0.97 \pm 0.10 \\
Direct & Local & 29.89 \pm 0.32  & 41.8 \pm 2.3 & 0.68 \pm 0.04 \\[1ex]
 & Local (w.a.) & 30.25 \pm 0.30 & 38.8 \pm 1.9 & 0.75 \pm 0.04 \\[3ex]
Staged & Nonlocal & 29.0 \pm 1.2 & 45 \pm 5 & 0.62 \pm 0.07 \\
Direct & Nonlocal & 29.2 \pm 2.0 & 38.2 \pm 2.9 & 0.73 \pm 0.07 \\[1ex]
 & Nonlocal (w.a.) & 29.1 \pm 1.0 & 39.9 \pm 2.5 & 0.70 \pm 0.05 \\
\botrule
\end{tabular*}

\end{table}

First,
we consider $\lambda_{L}$.
All of our $\lambda_{L}$s are considerably shorter than the often quoted
value of \qty{\sim 39}{\nano\meter}~\cite{1965-Maxfield-PR-139-A1515}
which is used extensively in technical applications of \ch{Nb}
(e.g., the modeling of \gls{srf} cavities).
Taking weighted averages~\footnote{Explicitly, the weighed average of a set of $n$ values $v_{i}$ with uncertainties $\delta v_{i}$ is given by: $ \left ( \sum_i^n w_{i} v_{i} \right ) /  \left ( \sum_i^n w_i \right )$, where the weights $w_i \equiv 1 / \delta v_i^2$. Similarly, the uncertainty in the weighted average is given by: $\sqrt{1 / \left ( \sum_i^n w_i \right )}$.}
of the values obtained from the local and nonlocal analyses separately,
we obtain a value in the range of \qtyrange{29}{30}{\nano\meter},
which is in excellent agreement with a recent literature average
(\qty{\sim 29}{\nano\meter})~\cite{2023-McFadden-PRA-19-044018,2023-McFadden-JAP-134-163902},
as well as several older \gls{le-musr}~\cite{2005-Suter-PRB-72-024506}
and
surface resistance~\cite{1999-Benvenuti-PC-316-153}
measurements.
The consistency between the local and nonlocal analyses is likely a consequence of $\lambda_{L}$'s 
close proximity to $\xi_{0}$ (see below),
making differences in the two $B(z)$ models subtle.
For comparison,
reports making use of a recent \gls{dft} calculation
of \ch{Nb}'s electronic structure~\cite{2023-Zarea-FP-11-1269872}
suggest
$\lambda_{L} \approx \qty{33}{\nano\meter}$~\cite{2022-Prozorov-PRB-106-L180505,2023-Zarea-FEM-3-1259401},
which is comparable in magnitude to our estimates,
though larger in absolute value by \qty{\sim 3}{\nano\meter}.
Overall,
we take the agreement of our $\lambda_{L}$s with other experiments,
along with their close proximity to the prediction from theory,
as a strong confirmation that we are probing
\ch{Nb}'s intrinsic value.

Next
we consider $\xi_{0}$,
whose averages from \Cref{tab:results} for the local and nonlocal cases
range from \qtyrange{38}{40}{\nano\meter}.
These values, derived from direct measurements,
are in excellent agreement with
the nominally quoted \qty{\sim 38}{\nano\meter}~\cite{1965-Maxfield-PR-139-A1515},
as well as a recent literature average
(\qty{\sim 40}{\nano\meter})~\cite{2023-McFadden-PRA-19-044018,2023-McFadden-JAP-134-163902}.
Such a level of agreement is encouraging,
as $\xi_{0}$ is the more challenging of the two length scales
to determine~\cite{2005-Suter-PRB-72-024506}.
Similar to $\lambda_{L}$,
$\xi_{0}$ can also be estimated from electronic structure
calculations~\cite{2023-Zarea-FP-11-1269872};
however,
the complexity of \ch{Nb}'s band structure adds ambiguity
to the comparison with experiment.
For example,
a multi-band average in Ref.~\citenum{2022-Prozorov-PRB-106-L180505} yields
$\xi_{0} \approx \qty{98}{\nano\meter}$,
whereas Ref.~\citenum{2023-Zarea-FEM-3-1259401} reports a single
value \qty{\sim 33}{\nano\meter}.
Both quantities differ from our experimental estimates,
though the latter is much closer in value
(difference of \qty{\sim 6}{\nano\meter}).
Despite the differences from theory,
we take the level of agreement with other experiments
as affirmation of the accuracy of our measurements.

Having verified the validity of our measurements,
we now consider the \emph{best} estimate for the two lengths.
The averaged results for the local and nonlocal limits in \Cref{tab:results}
show that the local approximation, despite neglecting nonlocal effects,
provides a good first-order estimate for the intrinsic length scales.
This is consistent with recent observations that the
local model provides a reasonable description of the Meissner response
in \gls{srf}-grade \ch{Nb}~\cite{2023-McFadden-PRA-19-044018};
however, the surfaces of such samples are often lightly doped.
Another more detailed analysis of Meissner screening in \ch{Nb}
has revealed some nonlocal character~\cite{2005-Suter-PRB-72-024506}.
Thus,
we take the average of the nonlocal results as our best estimate,
yielding
$\lambda_L = \qty{29.1 \pm 1.0}{\nano\meter}$
and
$\xi_0 = \qty{39.9 \pm 2.5}{\nano\meter}$.
With new values for $\lambda_L$ and $\xi_0$ established,
we now consider the implications of their revision.

On the fundamental side,
we comment on \ch{Nb}'s recently proposed ``intrinsic'' type-I superconductivity~\cite{2022-Prozorov-PRB-106-L180505}.
Such a proposition,
while surprising,
is appealing in that nearly all elemental superconductors
are type-I.
To assess this claim,
we calculate \ch{Nb}'s \gls{gl} parameter $\kappa$ using the well-known
Gor'kov expression
(see, e.g.,~\cite{1979-Orlando-PRB-19-4545}):
\begin{equation*}
	\kappa \approx 0.957 \frac{ \lambda_{L} }{ \xi_{0} } = \num{0.70 \pm 0.05} .
\end{equation*}
This value falls just below the \gls{gl} criterion for type-II superconductivity
(i.e., $\kappa > 1/\sqrt{2} \approx 0.707$),
suggesting that ``clean'' \ch{Nb} is a (borderline) type-I superconductor.
While the large uncertainty in $\kappa$ makes this claim tentative,
it provides some experimental backing to the notion of intrinsic type-I
behavior put forth in Ref.~\cite{2022-Prozorov-PRB-106-L180505}.
Additional support can be derived from magnetometry measurements on
ultra-pure \ch{Nb} samples (i.e., $\mathrm{RRR} \gtrsim \num{e4}$),
which show that the element is only of type-I over the narrow temperature range
$T_{c} - T < \qty{0.2}{\kelvin}$~\cite{1974-Alekeevskiy-FMM-37-63,arXiv:2506.01330}.
At lower temperatures,
``clean'' \ch{Nb} transitions into an intertype superconductor
(i.e., type-II/1),
displaying an attractive interaction of flux vortices over intermediate length scales
(see, e.g.,~\cite{1973-Auer-PRB-7-136,1987-Klein-JLTP-69-1,2017-Reimann-PRB-96-144506,2019-Backs-PRB-100-064503}).
The phase boundary between this intertype behavior and that of a ``classic'' type-II superconductor
(i.e., type-II/2) has been established
previously for \ch{Nb}~\cite{1987-Klein-JLTP-69-1,1987-Weber-JJAP-26-917,1989-Weber-PC-161-272}.
To be more conclusive about the type-I nature in close proximity to $T_c$,
direct measurement of the attractive nature of vortices is needed.
Efforts along this line are currently being made~\cite{2021-Ooi-PRB-104-064504,2025-Ooi-PRB-111-094519}.

On the practical side,
the updated $\xi_{0}$ has immediate consequences for maximizing the quality factor $Q$ in
\ch{Nb} \gls{srf} cavity resonators.
Following $Q$'s definition (i.e., the quotient of energy stored to power dissipated),
its value is inversely proportional to \ch{Nb}'s mean surface resistance $\bar{R}_{s}$,
which in the limit of weak Ohmic dissipation and local electrodynamics follows~\cite{2017-Gurevich-SST-30-034004}:
\begin{equation*}
	Q \propto \bar{R}_{s}^{-1} \propto \left [ \ell \left ( 1 + \frac{\pi}{2} \frac{ \xi_{0} }{ \ell } \right )^{3/2} \right ]^{-1} .
\end{equation*}
This quantity is maximized when $\ell = (\pi / 4) \xi_{0} = \qty{31.3 \pm 2.0}{\nano\meter}$.
Impurity levels matching this criteria are easily achieved using the oxygen-doping
treatments employed in this work
(particularly the steps used for sample \ch{Nb}-SR12)~\cite{2021-Lechner-APL-119-082601,2024-Lechner-JAP-135-133902}.


In summary,
we determined \ch{Nb}'s fundamental superconducting
length scales,
the London penetration depth $\lambda_{L}$
and
Pippard/\gls{bcs} coherence length $\xi_{0}$, 
using depth-resolved \gls{sims} and \gls{le-musr}
measurements on oxygen-doped \ch{Nb} samples,
prepared via contemporary \gls{srf} cavity methods.
We identify
$\lambda_{L} = \qty{29.1 \pm 1.0}{\nano\meter}$
and
$\xi_{0} = \qty{39.9 \pm 2.5}{\nano\meter}$,
consistent with other
authoritative
measurements
and
comparable to predictions from contemporary electronic structure calculations.
These
revised
lengths,
quantified simultaneously at the nanoscale,
imply
a \gls{gl} parameter of $\kappa = \num{0.70 \pm 0.05}$,
suggesting \ch{Nb} may be a type-I superconductor in the ultra-pure limit.
Additional experiments sensitive to the attractive nature of
vortices near \ch{Nb}'s $T_c$ will be necessary to confirm this claim.
Our updated
$\lambda_L$ and $\xi_0$
values
will be useful
in technical applications of \ch{Nb},
such as \gls{srf} cavities or engineered heterostructures,
where knowledge of the element's superconducting properties
are essential.

\begin{acknowledgments}
	T.J.\ acknowledges financial support from \acrshort{nserc}
	(grant SAPPJ-2024-00033).
	Additional support for E.M.L., C.E.R., and M.J.K.\ was provided by
	the U.S.\ Department of Energy, Office of Science,
	Office of Nuclear Physics under contract DE-AC05-06OR23177.
	The authors also gratefully acknowledge the Office of High Energy Physics,
	U.S.\ Department of Energy,
	for partial support of J.W.A.\ under Grant No.\ DE-SC-0018918 to Virginia Tech.
	This work is based on experiments performed at the
	\acrlong{sms},
	\acrlong{psi},
	Villigen, Switzerland
	(proposal number 20222354).
\end{acknowledgments}

\section*{Data Availability}

\Gls{le-musr} data that support the findings of this work are openly available~\cite{lem-data}.
All other data are available from the corresponding authors upon reasonable request.

\bibliography{references.bib,unpublished.bib}

\end{document}


\title{
	Supporting Material for:\\[1ex]
	Niobium's intrinsic coherence length and penetration depth revisited using low-energy muon spin spectroscopy
	and secondary-ion mass spectrometry
}


\author{Ryan~M.~L.~McFadden}
\email[E-mail: ]{rmlm@triumf.ca}
\affiliation{TRIUMF, 4004 Wesbrook Mall, Vancouver, BC V6T~2A3, Canada}
\affiliation{Department of Physics and Astronomy, University of Victoria, 3800 Finnerty Road, Victoria, BC V8P~5C2, Canada}

\author{Jonathan~W.~Angle}
\affiliation{Department of Materials Science and Engineering, Virginia Polytechnic Institute and State University, Blacksburg, Virginia 24061, USA}
\affiliation{Pacific Northwest National Laboratory, 902 Battelle Boulevard, Richland, WA 99354, United States of America}

\author{Eric~M.~Lechner}
\affiliation{Thomas Jefferson National Accelerator Facility, 12000 Jefferson Avenue, Newport News, Virginia 23606, United States of America}

\author{Michael~J.~Kelley}
\affiliation{Thomas Jefferson National Accelerator Facility, 12000 Jefferson Avenue, Newport News, Virginia 23606, United States of America}
\affiliation{Nanoscale Characterization and Fabrication Laboratory, Virginia Polytechnic Institute and State University, 1991 Kraft Drive, Blacksburg, Virginia 24061, USA}

\author{Charles~E.~Reece}
\affiliation{Thomas Jefferson National Accelerator Facility, 12000 Jefferson Avenue, Newport News, Virginia 23606, United States of America}

\author{Matthew~A.~Coble}
\affiliation{Pacific Northwest National Laboratory, 902 Battelle Boulevard, Richland, WA 99354, United States of America}

\author{Thomas~Prokscha}
\affiliation{PSI Center for Neutron and Muon Sciences CNM, Paul Scherrer Institute, Forschungsstrasse 111, 5232 Villigen, Switzerland}

\author{Zaher~Salman}
\affiliation{PSI Center for Neutron and Muon Sciences CNM, Paul Scherrer Institute, Forschungsstrasse 111, 5232 Villigen, Switzerland}

\author{Andreas~Suter}
\affiliation{PSI Center for Neutron and Muon Sciences CNM, Paul Scherrer Institute, Forschungsstrasse 111, 5232 Villigen, Switzerland}

\author{Tobias~Junginger}
\email[E-mail: ]{junginger@uvic.ca}
\affiliation{TRIUMF, 4004 Wesbrook Mall, Vancouver, BC V6T~2A3, Canada}
\affiliation{Department of Physics and Astronomy, University of Victoria, 3800 Finnerty Road, Victoria, BC V8P~5C2, Canada}

\date{\today}

\maketitle
\glsresetall
\vfill
\onecolumngrid
\pagebreak
%
\tableofcontents
\vfill
\listoftables
\vfill
\listoffigures
\vfill
\pagebreak
\twocolumngrid
\glsresetall

\section{
	Methods
	\label{sec:methods}
}

\subsection{
	Sample Preparation
	\label{sec:methods:samples}
}

\begin{table*}[p]
	\caption[Summary of specific surface and heat treatments used for each \ch{Nb} sample.]{
		\label{tab:samples}
		Summary of specific surface and heat treatments used for each \ch{Nb} sample.
		For further details see \Cref{sec:methods:samples}.
	}
	\begin{ruledtabular}
	\begin{tabular}{l l l l l}
		Sample & Step 6. | Anodization & Step 7. | Heat Treatment & Step 8. | Surface Removal & \acrshort{sims} Companion Sample \\
		\hline
		Nb-SR12 & \qty{25}{\volt} & \qty{300}{\celsius} for \qty{3}{\hour} & \qty{1}{\micro\meter} \gls{ep} &  Nb-SR3 / Nb-SR8 \\
		Nb-SR13 & --- & --- & --- & Nb-SR8 \\
		Nb-SR14 & --- & \qty{300}{\celsius} for \qty{3}{\hour} & \qty{300}{\nano\meter} \gls{ep} &  Nb-SR1 \\
		Nb-SR15 & \qty{5}{\volt} & \qty{300}{\celsius} for \qty{3}{\hour} & \qty{300}{\nano\meter} \gls{ep} &  Nb-SR2 \\
		Nb-SR16 & \qty{25}{\volt} & \qty{300}{\celsius} for \qty{3}{\hour} & \qty{300}{\nano\meter} \gls{ep} &  Nb-SR3 / Nb-SR8 \\
		Nb-SR18 & --- & \qty{350}{\celsius} for \qty{4}{\hour} & \qty{300}{\nano\meter} \gls{ep} &  Nb-SR4 \\
	\end{tabular}
	\end{ruledtabular}
\end{table*}

\ch{Nb} samples were prepared at Thomas Jefferson National Accelerator Facility
using high-purity and large \gls{rrr} (i.e., \num{> 300}) \ch{Nb} large grain stock sheets,
similar to those typically used for \gls{srf} cavity fabrication.
For consistency,
all samples were cut from the same grain within the stock,
with their orientation confirmed via electron backscatter diffraction.
For \gls{sims} measurements
(see \Cref{sec:methods:sims}),
``companion'' samples were used with the orientation of the surface marked
to ensure a common orientation was probed
(i.e., to avoid differences in \glspl{rsf} due to ion channeling
and
crystallographic orientation dependent sputtering~\cite{2022-Angle-JVSTB-40-024003}).
Following cutting of the samples
(\qty{\sim 2.5}{\centi\meter} diameter discs for
for \gls{le-musr}~\cite{2000-Morenzoni-PB-289-653,2008-Prokscha-NIMA-595-317}
and
\qtyproduct{6 x 10}{\milli\meter} plates for \gls{sims},
all with thicknesses of a few millimeters)
by wire electro-discharge machining,
each sample was prepared similarly to an \gls{srf} cavity.
Such treatments followed~\cite{2023-Lechner-PRAB-26-103101}:
\begin{enumerate}
	\item \qty{100}{\micro\meter} surface removal using \gls{bcp}.
	\item Mechanical polishing of the surface to an average roughness \qty{< 5}{\nano\meter}.
	\item \qty{20}{\micro\meter} surface removal using \gls{ep} (i.e., to remove surface damage from mechanical polishing).
	\item Vacuum annealing at \qty{800}{\celsius} for \qty{3}{\hour}.
	\item \qty{20}{\micro\meter} surface removal using \gls{ep} (i.e., to remove any surface contaminants accrued during heating).
	\item Surface anodization (optional).
	\item Heat treatment between \qtyrange{300}{350}{\celsius} (i.e., for surface oxide dissolution and oxygen doping).
	\item Surface removal by \gls{ep}
	(i.e., to remove other impurities that may have diffused in during the thermal treatment
	and tune the oxygen content).
\end{enumerate}
The preparation of all samples followed steps \numrange{1}{5} identically,
with sample-specific details for steps \numrange{6}{8} given in \Cref{tab:samples}.
Estimates for the concentrations of the major atomic impurities
carbon, nitrogen, and oxygen
(determined from \gls{sims} --- see \Cref{sec:methods:sims})
are given in Table~1 in the main text.

\subsection{
	\acrshort{sims}
	\label{sec:methods:sims}
}

\Gls{sims}
measurements were performed with a
CAMECA IMS 7f-GEO magnetic sector instrument.
The instrument was operated in negative ion mode using a \qty{70}{\nano\ampere}
\ch{Cs^{+}} primary beam at \qty{8}{\kilo\electronvolt} impact energy.
Both raster and crater sizes were \qtyproduct{150 x 150}{\micro\meter},
and a \qty{63}{\micro\meter} detection window was used to limit secondary ion collection
to the crater floor and minimize sidewall contributions.
No charge compensation was applied as the sample was conductive.
Secondary ions of \ch{^{12}C^{-}},
\ch{^{16}O^{-}},
\ch{^{93}Nb^{14}N^{-}},
and \ch{^{93}Nb^{-}} were detected,
with the corresponding impurity profiles for \ch{C}, \ch{N}, and \ch{O} reported here.
Data were converted to concentration units (ppma) and plotted against sputter depth in nanometers,
determined by using a Bruker stylus profilometer and calibrated sputter rates.
Initial quantification used a dedicated implant standard,
which was implanted with \qty{2e15}{atoms/\centi\meter\squared} of
\ch{^{12}C} (\qty{135}{\kilo\electronvolt}),
\ch{^{14}N} (\qty{160}{\kilo\electronvolt}),
and
\ch{^{18}O} (\qty{180}{\kilo\electronvolt}).
However, \ch{^{12}C} and \ch{^{14}N} suffered from high background signals that complicated \gls{rsf} calibration.
Additionally, the \ch{Nb^{-}} matrix signal varied unexpectedly across samples.
This variability violated key \gls{rsf} assumptions and led to inconsistent background concentrations across the dataset.

To address these challenges, the experimental samples themselves were implanted under the same conditions,
with \ch{^{13}C} used in place of \ch{^{12}C} to avoid overlap with ``background'' carbon
and to prevent interference with the \ch{^{12}C} profile.
By reusing \ch{^{18}O} as the implant species,
the background oxygen interference with the \ch{^{16}O} profile was likewise avoided.
Although \ch{^{15}N} implantation was considered, \ch{^{14}N} was retained due to cost constraints.
Thus, the nitrogen \gls{rsf} obtained in the implanted \gls{sims} analyses was then applied
retroactively to the prior analyses which lacked implant peaks, resulting in more robust quantitation.
This self-implantation strategy ensured that \glspl{rsf} were calculated using a perfectly matrix-matched substrate,
improving both accuracy and precision.
A forthcoming manuscript will provide a detailed account of the experimental strategies,
challenges, and calibration procedures undertaken to ensure the highest possible accuracy of these \gls{sims} measurements.
In the mean time,
additional technical details can be found in works describing measurements
on related samples~\cite{2021-Angle-JVSTB-39-024004,2021-Lechner-APL-119-082601,2022-Angle-JVSTB-40-024003}.

\subsection{
	\acrshort{le-musr}
	\label{sec:methods:le-musr}
}

\begin{figure}[p]
	\centering
	\includegraphics[width=1.0\columnwidth]{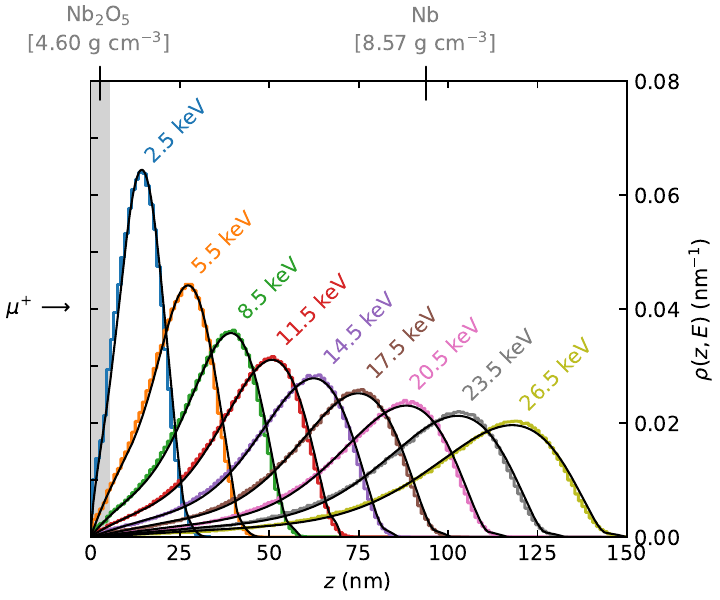}
	\caption[Typical stopping profiles for $\mu^{+}$ implanted in a \ch{Nb2O5}(\qty{5}{\nano\meter})/\ch{Nb} target at different energies.]{
		\label{fig:implantation-profiles}
		Typical stopping profiles $\rho(z, E)$ for $\mu^{+}$ implanted in a \ch{Nb2O5}(\qty{5}{\nano\meter})/\ch{Nb} target
		at different energies $E$ (indicated in the inset),
		simulated using the Monte Carlo code
		\texttt{TRIM.SP}~\cite{1991-Eckstein-SSMS-10,trimsp}.
		The profiles, represented here as histograms with \qty{1}{\nano\meter} bins,
		were generated from \num{e6} projectiles.
		Full simulation details can be found elsewhere~\cite{2002-Morenzoni-NIMB-192-245,2023-McFadden-PRA-19-044018,2026-McFadden-NIMB-570-165954}
		The solid black lines denote fits to an empirical model for
		the stopping distribution
		(see \Cref{sec:methods:staged}),
		whose parameters are listed in \Cref{tab:stopping-pars}.
	}
\end{figure}

\Gls{le-musr} measurements were performed at the \gls{sms},
located within the \gls{psi} in Villigen, Switzerland.
Using the $\mu E4$ beamline~\cite{2008-Prokscha-NIMA-595-317},
a \qty{\sim 100}{\percent} spin-polarized
\qty{\sim 15}{\kilo\electronvolt} $\mu^{+}$ beam with an intensity
of \qty{\sim e4}{\per\second}
was generated by moderating the energy of a
\qty{\sim 4}{\mega\electronvolt}  ``surface'' $\mu^{+}$ beam
using a film of condensed cryogenic gas~\cite{1994-Morenzoni-PRL-72-2793,2001-Prokscha-ASS-172-235}
and electrostatically re-accelerating the eluting epithermal (\qty{\sim 15}{\electronvolt})
muons.
The beam was delivered to a dedicated spectrometer~\cite{2000-Morenzoni-PB-289-653,2008-Prokscha-NIMA-595-317,2012-Salman-PP-30-55}
using electrostatic optics housed within an \gls{uhv} beamline,
with the $\mu^{+}$ arrival times triggered on a thin
(\qty{\sim 10}{\nano\meter}) carbon foil detector.
Note that passage through the foil results in both a slight reduction in
the beam's mean kinetic energy (\qty{\sim 1}{\kilo\electronvolt})
and introduces a small (asymmetric) energy spread (\qty{\sim 450}{\electronvolt}).
Control over the $\mu^{+}$ implantation energy
(and the $\mu^{+}$ stopping depth)
is achieved by biasing
an electrically isolated sample holder using a \gls{hv} power supply.
The stopping of $\mu^{+}$ in \ch{Nb} was simulated using the Monte Carlo code
\texttt{TRIM.SP}~\cite{1991-Eckstein-SSMS-10,trimsp},
providing spatial sensitivity to depths between
\qtyrange{\sim 10}{\sim 160}{\nano\meter}~\cite{2002-Morenzoni-NIMB-192-245,2023-McFadden-PRA-19-044018,2026-McFadden-NIMB-570-165954}.
Typical $\mu^{+}$ stopping profiles are shown in \Cref{fig:implantation-profiles}
(as well as the insets of \Cref{fig:staged-data-fits,fig:asymmetry-direct-local,fig:asymmetry-direct-nonlocal}
and Figure~1 in the main text).

In \gls{le-musr},
the implanted $\mu^{+}$ spins $\mathbf{S}$ reorient
in the local magnetic field $\mathbf{B}$
at their stopping site,
which is monitored
via the anisotropic $\beta$-emissions from $\mu^{+}$ decay
(mean lifetime $\tau_{\mu} = \qty{2.1969811 \pm 0.0000022}{\micro\second}$~\cite{2022-Workman-PTEP-2022-083C01-short}).
When $\mathbf{B}$ is transverse to the spin direction,
the expectation value $\langle S \rangle$ will precess at a rate equal to the probe's Larmor frequency:
\begin{equation}
	\label{eq:larmor}
	\omega_{\mu} = \gamma_{\mu} B ,
\end{equation}
where $\gamma_{\mu} / (2 \pi) = \qty{135.538 809 4 \pm 0.000 003 0}{\mega\hertz\per\tesla}$
is the muon gyromagnetic ratio~\cite{2021-Tiesinga-RMP-93-025010}.
In the experiments performed here,
this so-called transverse-field geometry was used
(see, e.g., Ref.~\cite{2024-Amato-IMSS})
wherein an external field $B_\mathrm{applied} \approx \qty{25}{\milli\tesla}$
was applied perpendicular to the initial direction of $\mu^{+}$ spin-polarization
and parallel to the surface of our \ch{Nb} samples.
This configuration is highly sensitive to inhomogeneities in $B$
and
allows for the local field distribution $p(B)$ to be determined,
which is directly related to \ch{Nb}'s Meissner screening profile $B(z)$.
Similar setups have been employed in related experiments~\cite{2005-Suter-PRB-72-024506,2014-Romanenko-APL-104-072601,2017-Junginger-SST-30-125013,2023-McFadden-PRA-19-044018}.

In \gls{le-musr},
the temporal evolution of the $\beta$-decay \emph{asymmetry} $A(t)$,
which is proportional to the spin-polarization of the $\mu^{+}$ ensemble $P_{\mu}(t) \equiv \langle S \rangle / S$,
is monitored in two
opposing detectors (denoted $+$ and $-$).
The count rate in a single detector $N_{\pm}$ is given by:
\begin{equation}
	\label{eq:counts}
	N_{\pm}(t) = N_{0, \pm} \exp \left ( - \frac{t}{\tau_{\mu}} \right ) \left [ 1 + A_{\pm}(t) \right ] + b_{\pm} ,
\end{equation}
where $N_{0, \pm}$ and $b_{\pm}$ are the incoming rates of
``good'' and ``background'' decay events,
and
$A_{\pm}(t) = A_{0,\pm} P_{\mu}(t)$.
For a \emph{perfect} detector pair
(i.e., identical detection efficiencies, mirrored geometry, etc.),
the latter component can be obtained directly from the experimental count rates
by forming their \emph{asymmetry}:
\begin{equation}
	\label{eq:asymmetry}
	A_{\mathrm{exp}}(t) = \frac{ \left [ N_{+}(t) - b_{+} \right ] - \left [ N_{-}(t) - b_{-} \right ] }{ \left [ N_{+}(t) - b_{+} \right ] + \left [ N_{-}(t) - b_{-} \right ] } ,
\end{equation}
where $A_{+}(t) = A_{-}(t) = A(t) \equiv A_{0} P_{\mu}(t)$,
with the proportionality constant $A_{0}$ typically in the range of \numrange{\sim 0.1}{\sim 0.3}.
In practice, this ideal situation is rarely realized
and
one typically obtains an asymmetry of the form:
\begin{equation}
	\label{eq:instrumental-asymmetry}
	A_{\mathrm{exp}}(t) = \frac{ \left  ( 1-\alpha \right ) +  \left ( 1 + \alpha \beta \right ) A_{0}P_{\mu}(t) }{ \left ( 1 + \alpha \right ) + \left ( 1 - \alpha \beta \right ) A_{0}P_{\mu}(t) } ,
\end{equation}
which includes the ``instrumental'' contribution from the parameters
$\alpha \equiv N_{0, -} / N_{0, +}$ and $\beta \equiv A_{0, -} / A_{0, +}$
being different from unity
(see, e.g., Ref.~\cite{2024-Amato-IMSS})
While \Cref{eq:instrumental-asymmetry} can be used directly in fitting
(i.e., by including $\alpha$ and $\beta$ as fit parameters)
it is often easier to fit each detector's signal independently with \Cref{eq:counts}.
In either instance,
the ``corrected'' asymmetry is obtained by inverting the relationship in \Cref{eq:instrumental-asymmetry}:
\begin{equation}
	\label{eq:corrected-asymmetry}
	A_{0} P_{\mu}(t) = \frac{ \left ( \alpha - 1 \right ) \left ( \alpha + 1 \right ) A_{\mathrm{exp}}(t)}{ \left ( \alpha \beta + 1 \right ) \left ( \alpha\beta - 1 \right ) A_{\mathrm{exp}}(t)} \equiv {A}(t).
\end{equation}
In this work,
we analyze each $N_{\pm}(t)$ independently using \Cref{eq:counts},
with the ``corrected'' asymmetry $A(t)$ subsequently derived from \Cref{eq:asymmetry,eq:corrected-asymmetry}
(i.e., for plotting).

In a \gls{le-musr} experiment,
the physics of the material under study is encapsulated entirely by $P_{\mu}(t)$,
which determines the time-dependence of $A(t)$.
In a transverse field geometry,
$P_{\mu}(t)$ depends on the local field distribution $p(B)$ according to:
\begin{equation}
	\label{eq:polarization}
	P_{\mu}(t) = \int_{0}^{+\infty} p(B) \cos ( \omega_{\mu} t + \phi_{\pm} ) \, \mathrm{d}B ,
\end{equation}
where $t$ is the time (in \unit{\micro\second}) after $\mu^{+}$ implantation,
$\omega_{\mu}$ is given by \Cref{eq:larmor},
and
$\phi_{\pm}$ is a (detector-dependent) phase factor that depends on both
the $\mu^{+}$ spin-orientation before implantation
and
the instrumental setup.

\subsection{
	\label{sec:methods:staged}
	``Staged'' Meissner Screening Analysis
}

For \ch{Nb} in its normal state at temperatures $T \geq T_{c}$,
its internal $p(B)$ is,
to an excellent approximation,
given by a Gaussian distribution,
with \Cref{eq:polarization} becoming:
\begin{equation}
	\label{eq:polarization-gaussian}
	P_{\mu}(t) = \exp \left ( - \frac{\sigma^{2} t^{2}}{2} \right ) \cos \left ( \gamma_{\mu} \langle B \rangle t + \phi_{\pm} \right ) ,
\end{equation}
where $\sigma$ is a damping parameter (related the distribution's width)
and
$\langle B \rangle$ is the mean of $p(B)$
(i.e., the Gaussian ``location'' parameter).
In the Meissner state
(i.e., at $T < T_{c}$ and $B < B_{c1}$),
flux expulsion from the element's interior
causes $p(B)$ to broaden asymmetrically.
The extent of this skewing depends on the depth region sampled
by the implanted $\mu^{+}$.
Phenomenologically,
this behavior can be described by a skewed Gaussian
distribution~\cite{2023-McFadden-PRA-19-044018,2024-Asaduzzaman-SST-37-025002}
or,
more generally,
using a sum of Gaussians:
\begin{equation}
	\label{eq:polarization-gaussian-sum}
	P_{\mu}(t) = \sum_{i=1}^{n} \eta_{i} \exp \left ( -\frac{\sigma_{i}^{2}t^{2}}{2} \right ) \cos \left ( \gamma_{\mu} \langle B_{i} \rangle t + \phi_{\pm} \right ) ,
\end{equation}
where $\eta_{i}$, $\langle B_j \rangle$, and $\sigma_i$ denote the
amplitude, mean, and damping rate
of the $i$\textsuperscript{th} component.
Empirically,
we find that the sum in \Cref{eq:polarization-gaussian-sum} can be truncated at $n \leq 3$,
but note that all three terms are not needed in every instance.
An implementation of
\Cref{eq:polarization-gaussian,eq:polarization-gaussian-sum}
is available in \texttt{musrfit}~\cite{2012-Suter-PP-30-69},
which we use here for our analysis.

\begin{figure*}[p]
\centering
\newcommand{\customwidth}{0.31\textwidth}
\includegraphics[width=\customwidth]{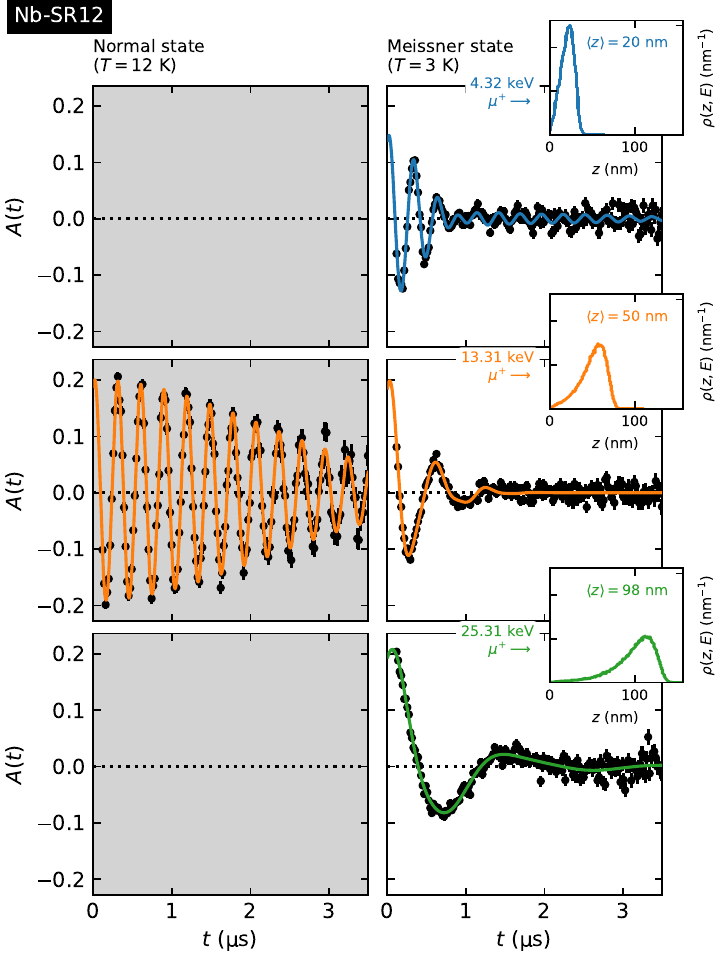}
\hfill
\includegraphics[width=\customwidth]{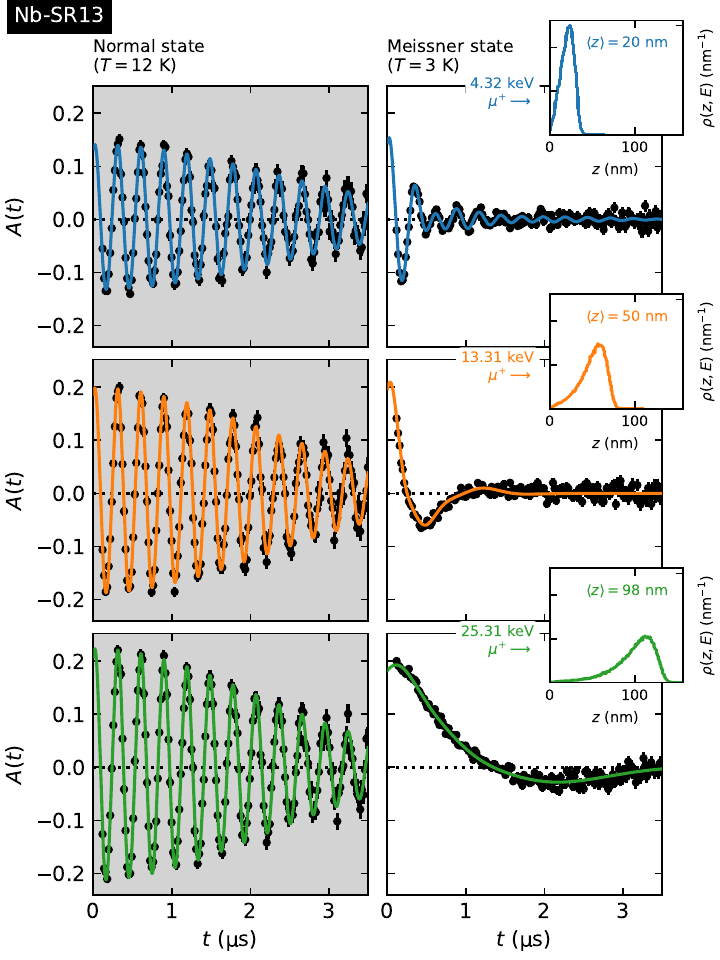}
\hfill
\includegraphics[width=\customwidth]{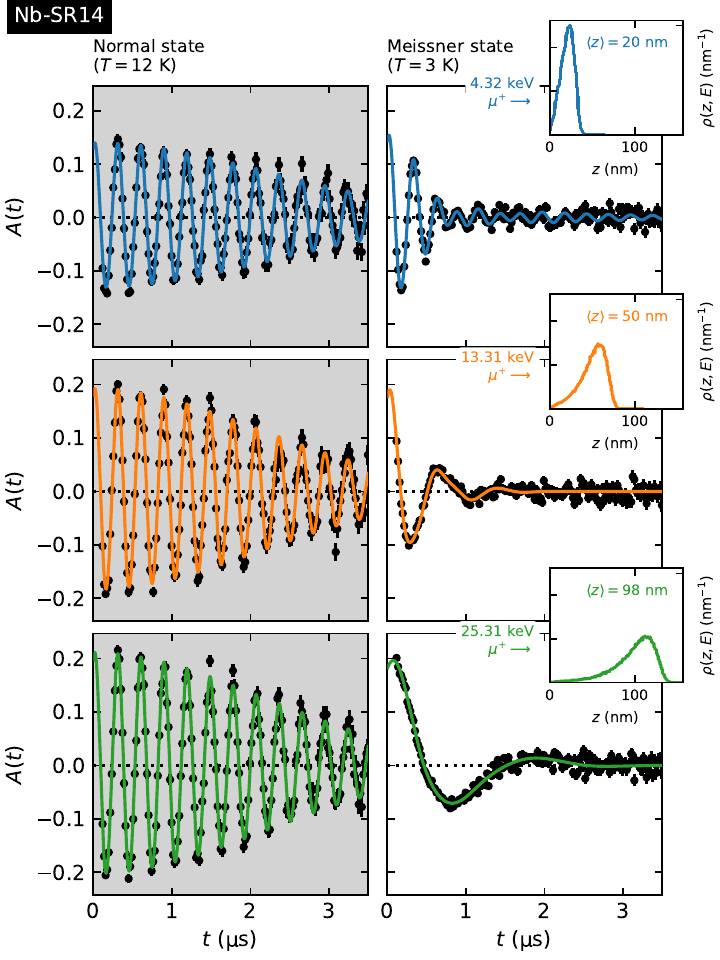} \\
\includegraphics[width=\customwidth]{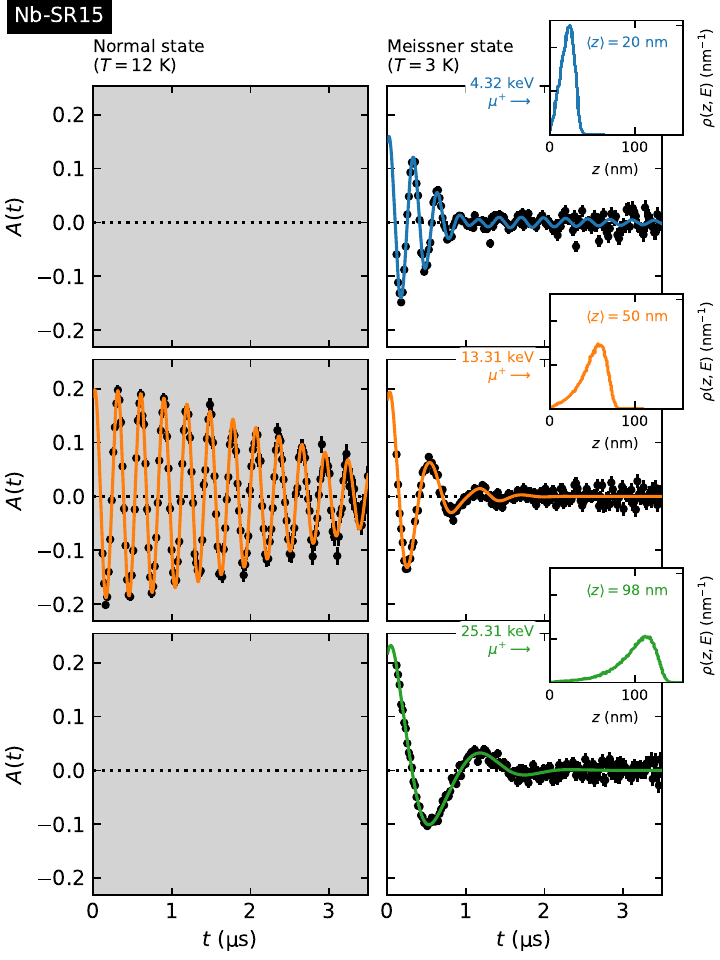}
\qquad
\includegraphics[width=\customwidth]{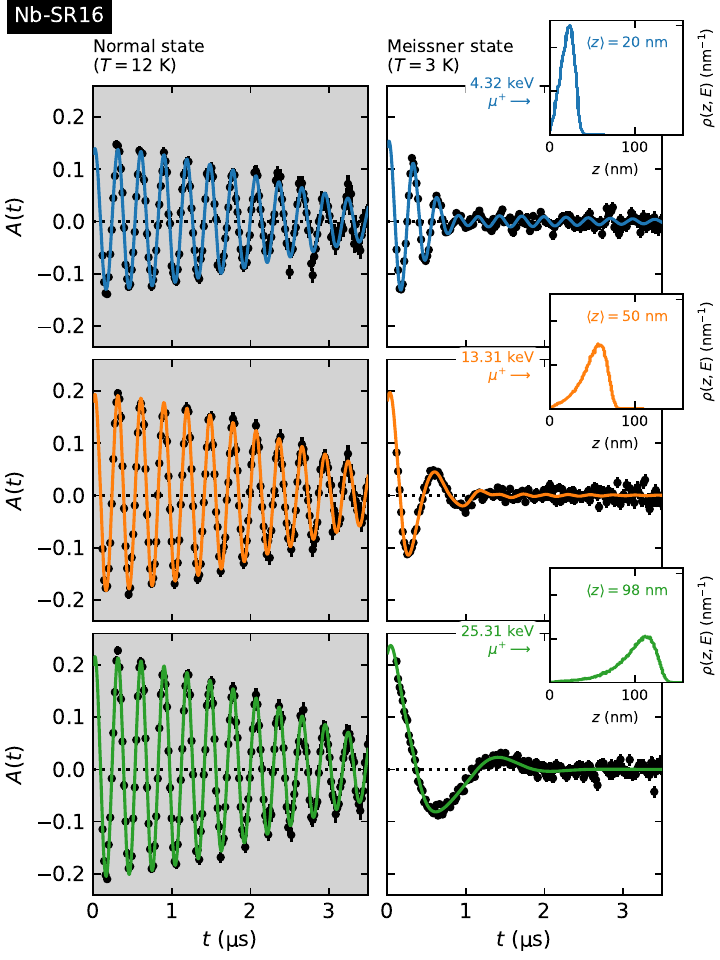}
\caption[
	\acrshort{le-musr} data in the \ch{Nb} samples fit using the ``staged'' approach.
]{
	\label{fig:staged-data-fits}
	Typical \gls{le-musr} data in several \ch{Nb} samples
	(\ch{Nb}-SR12, \ch{Nb}-SR13, \ch{Nb}-SR14, \ch{Nb}-SR15, and \ch{Nb}-SR16).
	In the normal state,
	the $\mu^{+}$ asymmetry $A(t)$ is weakly damped
	with a spin-precession rate that is independent of
	implantation energy $E$.
	By contrast,
	the damping of $A(t)$ in the Meissner state
	is strong, increasing with increasing $E$,
	which is accompanied by a decrease in the
	rate of spin-precession.
	The solid colored lines denote fits to the ``staged'' analysis model,
	wherein the field distribution is approximated
	as either a single Gaussian
	[\Cref{eq:polarization-gaussian}, normal state data]
	or
	a sum of Gaussians
	[\Cref{eq:polarization-gaussian-sum}, Meissner state data].
	The $\mu^{+}$ implantation profile $\rho(z, E)$
	and
	mean stopping depth $\langle z \rangle$,
	simulated using the
	\texttt{TRIM.SP} Monte Carlo code~\cite{1991-Eckstein-SSMS-10,trimsp},
	are shown in the inset for each $E$.
	Data and fits for sample \ch{Nb}-SR18 are shown in Figure~1 in the main text.
}
\end{figure*}

Note that,
in addition to the comments above,
we also imposed constraints on the detector phases
$\phi_{\pm}$ in \Cref{eq:counts}
to improve the robustness of the fitting procedure.
That is,
each $\phi_{\pm}$ was shared as a common fit parameter
for all measurements in each sample's dataset.
Crucially,
this ``global'' fitting procedure prevents some of the fit parameters
from converging to non-physical values,
but without penalty to the overall goodness-of-fit
(see, e.g., Refs.~\citenum{2023-McFadden-PRA-19-044018,2024-McFadden-APL-124-086101}).
This is easily achieved using \texttt{musrfit}~\cite{2012-Suter-PP-30-69}.
A subset of the fits obtained using this procedure
are shown in \Cref{fig:staged-data-fits}
and Figure~1 in the main text.

To extract the Meisner profile $B(z)$ from the \gls{le-musr} data,
for each measurement
(i.e., each $\mu^{+}$ implantation energy $E$)
we identify the corresponding mean magnetic field $\langle B \rangle$
using (see, e.g.,~\cite{2024-Amato-IMSS}):
\begin{equation}
	\label{eq:mean-gaussian-sum}
	\langle B \rangle (E) = \frac{1}{\tilde{\eta}(E)} \sum_{j=i}^{n} \eta_i(E) \cdot \langle B_i \rangle (E) ,
\end{equation}
where
\begin{equation*}
	\tilde{\eta}(E) \equiv \sum_{j=i}^{n} \eta_i (E) ,
\end{equation*}
and the $E$-dependence of the measured parameters has been made explicit.
It follows that $\langle B \rangle$ is related to the \emph{true} screening profile $B(z)$
through Equation~(4) in the main text,
which requires knowledge of the
$\mu^{+}$ stopping profile $\rho(z, E)$
(i.e., the kernel of the integral transform).
As $\rho(z, E)$ is usually obtained from Monte Carlo simulations
at select $E$s,
we note that its discrete nature isn't amenable for the direct evaluation of
$\langle B \rangle (E)$ as a smooth, continuous function.
To overcome this limitation,
we follow the approach described in related studies~\cite{2023-McFadden-PRA-19-044018,2024-Asaduzzaman-SST-37-025002}
and make use of an empirical model to \emph{analytically} define $\rho(z,E)$
over the range of $\mu^{+}$ implantation energies used here.

In general,
$\rho(z, E)$ can be expressed as a weighted sum of probability distributions:
\begin{equation}
	\label{eq:stopping-general}
	\rho(z, E) = \sum_{i}^{n} f_{i} p_{i} (z) ,
\end{equation}
where
$p_{i}(z)$ is a probability density function,
$z$ is the depth below the surface,
and
$f_{i} \in [0, 1]$ is the $i^{\mathrm{th}}$ weighting factor
obeying the constraint:
\begin{equation*}
	\sum_{i}^{n} f_{i} \equiv 1 .
\end{equation*}
For a \ch{Nb2O5}(\qty{5}{\nano\meter})/\ch{Nb} target,
where the thin oxide layer accounts for the surface's natural
oxidation~\cite{1987-Halbritter-APA-43-1},
the sum in \Cref{eq:stopping-general} can be truncated at
$n = 2$~\cite{2023-McFadden-PRA-19-044018}
and
rewritten as:
\begin{equation}
	\label{eq:stopping}
	\rho(z, E) = f_{1} p_{1}(z) + (1 - f_{1}) p_{2}(z) ,
\end{equation}
requiring only a single weight $f_{1}$.
In \Cref{eq:stopping},
the $p_{i}(z)$s are given by a modified beta distribution,
defined as:
\begin{equation}
\label{eq:beta-pdf}
p_{i}(z) =
	\begin{cases}
		\dfrac{ \left ( \frac{z}{z_{\mathrm{max},i}} \right )^{\alpha_i -1}  \left (1 -  \frac{z}{z_{\mathrm{max},i}} \right )^{\beta_i - 1}  }{ z_{\mathrm{max},i} \cdot B ( \alpha_i, \beta_i ) } , & 0 \leq z \leq z_{\mathrm{max}, i} , \\
		0, & \mathrm{elsewhere} ,
	\end{cases}
\end{equation}
where $z \in [0, z_{\mathrm{max},i}]$ is the depth below the surface,
$B ( a, b )$ denotes the beta function:
\begin{equation*}
B ( a, b ) \equiv \frac{ \Gamma (a) \Gamma (b) }{ \Gamma (a + b) } ,
\end{equation*}
and $\Gamma (s)$ is the gamma function:
\begin{equation*}
\Gamma (s) \equiv \int_{0}^{\infty} x^{s-1} \exp (-x) \, \mathrm{d}x .
\end{equation*}
Note that \Cref{eq:beta-pdf} differs from the usual beta distribution
in that its domain has been mapped from $[0, 1] \rightarrow [0, z_{\mathrm{max},i}]$,
with the ``extra'' $z_{\mathrm{max},i}$ in the expression's denominator
ensuring dimensional correctness
and
fulfillment of the normalization condition:
\begin{equation*}
\int_{0}^{\infty} p_i(z) \, \mathrm{d}z = 1 .
\end{equation*}

\begin{table*}[p]
	\caption[Empirical parameters describing $\mu^{+}$ stopping in \ch{Nb2O5}(\qty{5}{\nano\meter})/\ch{Nb}.]{
		\label{tab:stopping-pars}
		Empirical parameters describing $\mu^{+}$ stopping in \ch{Nb2O5}(\qty{5}{\nano\meter})/\ch{Nb}.
		The parameters were obtained for fits of \Cref{eq:stopping,eq:beta-pdf}
		to $\mu^{+}$ stopping profiles in \ch{Nb2O5}(\qty{5}{\nano\meter})/\ch{Nb}
		[mass densities
		\qty{4.60}{\gram\per\centi\meter\cubed}
		and
		\qty{8.57}{\gram\per\centi\meter\cubed}
		for the \ch{Nb2O5} and \ch{Nb} layers,
		respectively]
		simulated using the Monte Carlo code
		\texttt{TRIM.SP}~\cite{1991-Eckstein-SSMS-10,trimsp}.
		Here,
		$E$ is the muon implantation energy,
		$\alpha_{i}$,
		$\beta_{i}$,
		and $z_{\mathrm{max},i}$ denote the shape shape parameters in \Cref{eq:beta-pdf},
		and 
		$f_{1}$ is the weighting factor in \Cref{eq:stopping}.
		As described in \Cref{sec:methods:staged},
		using (linearly) interpolated values for these quantities
		as input for \Cref{eq:stopping,eq:beta-pdf} facilitates the generation
		of $\rho(z, E)$ at arbitrary implantation energy within
		$\qty{0.5}{\kilo\electronvolt} \leq E \leq \qty{30}{\kilo\electronvolt}$.
		Similar approaches have been used elsewhere~\cite{2023-McFadden-PRA-19-044018,2024-Asaduzzaman-SST-37-025002}.
	}
	\begin{tabular*}{\textwidth}{S @{\extracolsep{\fill}} S S S S S S S}
\botrule
{$E$ (\unit{\kilo\electronvolt})} & {$\alpha_{1}$} & {$\beta_{1}$} & {$z_{\mathrm{max},1}$ (\unit{\nano\meter})} & {$f_{1}$} & {$\alpha_{2}$} & {$\beta_{2}$} & {$z_{\mathrm{max},2}$ (\unit{\nano\meter})} \\
\hline
0.5 & 3.47 \pm 0.06 & 13.8 \pm 0.6 & 32.5 \pm 0.8 & 0.828 \pm 0.004 & 1.736 \pm 0.020 & 3.01 \pm 0.10 & 4.41 \pm 0.08 \\
1.0 & 3.381 \pm 0.023 & 8.56 \pm 0.13 & 29.31 \pm 0.20 & 0.9195 \pm 0.0012 & 1.797 \pm 0.022 & 2.49 \pm 0.07 & 4.21 \pm 0.04 \\
1.5 & 4.67 \pm 0.05 & 9.20 \pm 0.25 & 33.4 \pm 0.4 & 0.598 \pm 0.010 & 1.683 \pm 0.009 & 3.02 \pm 0.06 & 22.29 \pm 0.08 \\
2.0 & 4.646 \pm 0.034 & 7.08 \pm 0.08 & 33.90 \pm 0.15 & 0.582 \pm 0.005 & 1.695 \pm 0.010 & 2.234 \pm 0.025 & 23.054 \pm 0.027 \\
2.5 & 5.00 \pm 0.04 & 6.52 \pm 0.07 & 36.04 \pm 0.12 & 0.555 \pm 0.005 & 1.764 \pm 0.010 & 2.144 \pm 0.020 & 25.992 \pm 0.024 \\
3.0 & 5.10 \pm 0.04 & 5.82 \pm 0.05 & 37.98 \pm 0.08 & 0.530 \pm 0.004 & 1.769 \pm 0.010 & 1.891 \pm 0.017 & 27.929 \pm 0.017 \\
3.5 & 5.86 \pm 0.05 & 6.12 \pm 0.05 & 40.96 \pm 0.09 & 0.501 \pm 0.004 & 1.863 \pm 0.009 & 1.985 \pm 0.016 & 30.964 \pm 0.019 \\
4.0 & 6.58 \pm 0.05 & 6.33 \pm 0.05 & 43.70 \pm 0.07 & 0.482 \pm 0.004 & 1.952 \pm 0.008 & 2.081 \pm 0.015 & 33.975 \pm 0.023 \\
4.5 & 6.45 \pm 0.05 & 5.66 \pm 0.06 & 45.21 \pm 0.10 & 0.493 \pm 0.004 & 1.944 \pm 0.008 & 1.962 \pm 0.015 & 35.938 \pm 0.020 \\
5.0 & 6.62 \pm 0.05 & 5.40 \pm 0.05 & 47.34 \pm 0.09 & 0.494 \pm 0.004 & 1.955 \pm 0.008 & 1.901 \pm 0.015 & 37.933 \pm 0.019 \\
5.5 & 7.18 \pm 0.05 & 5.75 \pm 0.06 & 50.21 \pm 0.10 & 0.488 \pm 0.004 & 1.979 \pm 0.008 & 1.918 \pm 0.013 & 40.925 \pm 0.020 \\
6.0 & 7.21 \pm 0.05 & 5.46 \pm 0.05 & 52.23 \pm 0.09 & 0.489 \pm 0.004 & 1.999 \pm 0.008 & 1.857 \pm 0.013 & 42.901 \pm 0.018 \\
6.5 & 7.29 \pm 0.05 & 5.29 \pm 0.04 & 54.28 \pm 0.08 & 0.4979 \pm 0.0035 & 1.980 \pm 0.008 & 1.786 \pm 0.012 & 44.870 \pm 0.017 \\
7.0 & 7.95 \pm 0.05 & 5.57 \pm 0.04 & 56.70 \pm 0.08 & 0.523 \pm 0.004 & 2.070 \pm 0.008 & 2.059 \pm 0.014 & 48.848 \pm 0.029 \\
7.5 & 7.60 \pm 0.05 & 5.24 \pm 0.04 & 59.02 \pm 0.07 & 0.4955 \pm 0.0032 & 1.989 \pm 0.008 & 1.665 \pm 0.011 & 48.839 \pm 0.014 \\
8.0 & 7.37 \pm 0.05 & 4.79 \pm 0.04 & 60.39 \pm 0.08 & 0.5230 \pm 0.0033 & 1.970 \pm 0.008 & 1.642 \pm 0.011 & 50.821 \pm 0.014 \\
8.5 & 5.77 \pm 0.04 & 2.813 \pm 0.017 & 55.30 \pm 0.04 & 0.589 \pm 0.004 & 1.934 \pm 0.009 & 1.954 \pm 0.009 & 58.908 \pm 0.024 \\
9.0 & 5.78 \pm 0.05 & 2.660 \pm 0.017 & 56.168 \pm 0.035 & 0.542 \pm 0.004 & 1.881 \pm 0.010 & 1.571 \pm 0.007 & 58.837 \pm 0.014 \\
9.5 & 6.03 \pm 0.04 & 2.782 \pm 0.017 & 59.16 \pm 0.04 & 0.568 \pm 0.004 & 1.920 \pm 0.009 & 1.709 \pm 0.008 & 61.903 \pm 0.019 \\
10.0 & 5.80 \pm 0.04 & 2.556 \pm 0.016 & 60.131 \pm 0.032 & 0.556 \pm 0.004 & 1.856 \pm 0.010 & 1.467 \pm 0.007 & 62.767 \pm 0.011 \\
10.5 & 6.09 \pm 0.04 & 2.696 \pm 0.016 & 63.16 \pm 0.04 & 0.580 \pm 0.004 & 1.902 \pm 0.009 & 1.599 \pm 0.007 & 65.836 \pm 0.016 \\
11.0 & 6.11 \pm 0.04 & 2.650 \pm 0.015 & 65.15 \pm 0.04 & 0.585 \pm 0.004 & 1.887 \pm 0.009 & 1.541 \pm 0.007 & 67.795 \pm 0.014 \\
11.5 & 6.08 \pm 0.04 & 2.590 \pm 0.015 & 67.172 \pm 0.034 & 0.592 \pm 0.004 & 1.878 \pm 0.010 & 1.500 \pm 0.007 & 69.782 \pm 0.013 \\
12.0 & 6.42 \pm 0.04 & 2.726 \pm 0.015 & 70.08 \pm 0.04 & 0.6062 \pm 0.0034 & 1.945 \pm 0.009 & 1.626 \pm 0.007 & 72.843 \pm 0.018 \\
12.5 & 6.41 \pm 0.04 & 2.670 \pm 0.014 & 72.11 \pm 0.04 & 0.6144 \pm 0.0033 & 1.926 \pm 0.010 & 1.572 \pm 0.007 & 74.814 \pm 0.016 \\
13.0 & 6.75 \pm 0.04 & 2.810 \pm 0.016 & 75.05 \pm 0.05 & 0.6271 \pm 0.0031 & 1.979 \pm 0.009 & 1.703 \pm 0.008 & 77.895 \pm 0.023 \\
13.5 & 6.68 \pm 0.04 & 2.743 \pm 0.015 & 77.02 \pm 0.05 & 0.6350 \pm 0.0031 & 1.950 \pm 0.009 & 1.626 \pm 0.008 & 79.845 \pm 0.020 \\
14.0 & 6.68 \pm 0.04 & 2.685 \pm 0.013 & 79.24 \pm 0.04 & 0.6498 \pm 0.0029 & 1.98 \pm 0.01 & 1.704 \pm 0.008 & 82.812 \pm 0.021 \\
14.5 & 6.85 \pm 0.04 & 2.778 \pm 0.014 & 82.20 \pm 0.04 & 0.6733 \pm 0.0027 & 1.999 \pm 0.010 & 1.813 \pm 0.009 & 85.848 \pm 0.027 \\
15.0 & 6.94 \pm 0.04 & 2.739 \pm 0.013 & 84.18 \pm 0.04 & 0.6670 \pm 0.0027 & 2.013 \pm 0.010 & 1.757 \pm 0.008 & 87.860 \pm 0.026 \\
15.5 & 7.04 \pm 0.04 & 2.808 \pm 0.014 & 87.11 \pm 0.05 & 0.6873 \pm 0.0027 & 2.023 \pm 0.010 & 1.839 \pm 0.009 & 90.814 \pm 0.029 \\
16.0 & 7.10 \pm 0.04 & 2.764 \pm 0.013 & 89.15 \pm 0.05 & 0.6847 \pm 0.0026 & 2.015 \pm 0.010 & 1.765 \pm 0.008 & 92.832 \pm 0.027 \\
16.5 & 7.28 \pm 0.04 & 2.824 \pm 0.017 & 91.89 \pm 0.07 & 0.6988 \pm 0.0025 & 2.048 \pm 0.010 & 1.861 \pm 0.011 & 95.820 \pm 0.033 \\
17.0 & 7.33 \pm 0.04 & 2.798 \pm 0.013 & 94.21 \pm 0.04 & 0.7070 \pm 0.0024 & 2.051 \pm 0.010 & 1.896 \pm 0.009 & 98.753 \pm 0.029 \\
17.5 & 7.29 \pm 0.04 & 2.731 \pm 0.012 & 96.27 \pm 0.04 & 0.7034 \pm 0.0024 & 2.061 \pm 0.010 & 1.821 \pm 0.008 & 100.760 \pm 0.027 \\
18.0 & 7.52 \pm 0.04 & 2.799 \pm 0.013 & 99.11 \pm 0.05 & 0.7125 \pm 0.0024 & 2.102 \pm 0.010 & 1.902 \pm 0.009 & 103.771 \pm 0.032 \\
18.5 & 7.61 \pm 0.04 & 2.844 \pm 0.015 & 101.94 \pm 0.07 & 0.7291 \pm 0.0023 & 2.093 \pm 0.010 & 1.968 \pm 0.011 & 106.77 \pm 0.04 \\
19.0 & 7.589 \pm 0.035 & 2.775 \pm 0.012 & 104.05 \pm 0.05 & 0.7266 \pm 0.0022 & 2.071 \pm 0.010 & 1.857 \pm 0.009 & 108.744 \pm 0.030 \\
19.5 & 7.69 \pm 0.04 & 2.804 \pm 0.016 & 106.83 \pm 0.08 & 0.7359 \pm 0.0022 & 2.082 \pm 0.010 & 1.913 \pm 0.011 & 111.720 \pm 0.033 \\
20.0 & 7.79 \pm 0.04 & 2.781 \pm 0.012 & 109.16 \pm 0.05 & 0.7311 \pm 0.0022 & 2.091 \pm 0.010 & 1.848 \pm 0.009 & 113.755 \pm 0.031 \\
20.5 & 7.672 \pm 0.035 & 2.684 \pm 0.011 & 111.29 \pm 0.04 & 0.7366 \pm 0.0022 & 2.098 \pm 0.010 & 1.841 \pm 0.008 & 116.680 \pm 0.025 \\
21.0 & 7.83 \pm 0.04 & 2.734 \pm 0.011 & 114.25 \pm 0.04 & 0.7457 \pm 0.0021 & 2.115 \pm 0.010 & 1.898 \pm 0.009 & 119.638 \pm 0.025 \\
21.5 & 7.763 \pm 0.035 & 2.640 \pm 0.011 & 116.24 \pm 0.04 & 0.7391 \pm 0.0021 & 2.094 \pm 0.010 & 1.780 \pm 0.008 & 121.686 \pm 0.024 \\
22.0 & 7.92 \pm 0.04 & 2.675 \pm 0.011 & 119.05 \pm 0.05 & 0.7365 \pm 0.0022 & 2.094 \pm 0.010 & 1.730 \pm 0.008 & 123.711 \pm 0.026 \\
22.5 & 7.89 \pm 0.04 & 2.595 \pm 0.010 & 121.11 \pm 0.04 & 0.7286 \pm 0.0022 & 2.071 \pm 0.010 & 1.619 \pm 0.007 & 125.713 \pm 0.022 \\
23.0 & 7.964 \pm 0.035 & 2.627 \pm 0.010 & 124.28 \pm 0.04 & 0.7478 \pm 0.0020 & 2.094 \pm 0.010 & 1.739 \pm 0.008 & 129.672 \pm 0.024 \\
23.5 & 8.14 \pm 0.04 & 2.659 \pm 0.010 & 127.15 \pm 0.04 & 0.7504 \pm 0.0020 & 2.114 \pm 0.010 & 1.765 \pm 0.008 & 132.646 \pm 0.023 \\
24.0 & 8.10 \pm 0.04 & 2.572 \pm 0.010 & 129.16 \pm 0.04 & 0.7343 \pm 0.0021 & 2.063 \pm 0.010 & 1.566 \pm 0.007 & 133.685 \pm 0.020 \\
24.5 & 8.147 \pm 0.035 & 2.589 \pm 0.010 & 132.23 \pm 0.04 & 0.7502 \pm 0.0020 & 2.10 \pm 0.01 & 1.674 \pm 0.007 & 137.672 \pm 0.022 \\
25.0 & 8.46 \pm 0.04 & 2.705 \pm 0.012 & 135.90 \pm 0.06 & 0.7613 \pm 0.0020 & 2.172 \pm 0.011 & 1.830 \pm 0.009 & 141.634 \pm 0.026 \\
25.5 & 8.23 \pm 0.04 & 2.515 \pm 0.010 & 137.15 \pm 0.04 & 0.7370 \pm 0.0021 & 2.071 \pm 0.010 & 1.498 \pm 0.007 & 141.712 \pm 0.020 \\
26.0 & 8.26 \pm 0.04 & 2.521 \pm 0.009 & 140.24 \pm 0.04 & 0.7508 \pm 0.0020 & 2.089 \pm 0.011 & 1.581 \pm 0.007 & 145.648 \pm 0.019 \\
26.5 & 8.53 \pm 0.04 & 2.625 \pm 0.010 & 144.01 \pm 0.05 & 0.7617 \pm 0.0019 & 2.143 \pm 0.011 & 1.712 \pm 0.008 & 149.662 \pm 0.024 \\
27.0 & 8.63 \pm 0.04 & 2.636 \pm 0.010 & 147.00 \pm 0.05 & 0.7657 \pm 0.0019 & 2.155 \pm 0.011 & 1.714 \pm 0.008 & 152.671 \pm 0.026 \\
27.5 & 8.47 \pm 0.04 & 2.521 \pm 0.009 & 149.18 \pm 0.04 & 0.757 \pm 0.002 & 2.119 \pm 0.011 & 1.594 \pm 0.007 & 154.675 \pm 0.022 \\
28.0 & 8.62 \pm 0.04 & 2.536 \pm 0.009 & 152.16 \pm 0.04 & 0.7567 \pm 0.0019 & 2.126 \pm 0.011 & 1.579 \pm 0.007 & 157.656 \pm 0.020 \\
28.5 & 8.60 \pm 0.04 & 2.516 \pm 0.009 & 155.19 \pm 0.04 & 0.7621 \pm 0.0019 & 2.126 \pm 0.011 & 1.576 \pm 0.007 & 160.687 \pm 0.023 \\
29.0 & 8.75 \pm 0.04 & 2.522 \pm 0.009 & 158.03 \pm 0.04 & 0.7516 \pm 0.0020 & 2.123 \pm 0.011 & 1.497 \pm 0.007 & 162.684 \pm 0.020 \\
29.5 & 8.77 \pm 0.04 & 2.507 \pm 0.009 & 161.00 \pm 0.04 & 0.7551 \pm 0.0020 & 2.117 \pm 0.011 & 1.486 \pm 0.007 & 165.693 \pm 0.021 \\
30.0 & 8.90 \pm 0.04 & 2.507 \pm 0.009 & 163.94 \pm 0.05 & 0.751 \pm 0.002 & 2.124 \pm 0.011 & 1.457 \pm 0.007 & 168.692 \pm 0.020 \\
\botrule
\end{tabular*}
\end{table*}

For good ``coverage'' of $\rho(z, E)$ across the range of $E$s achievable by \gls{le-musr},
$\mu^{+}$ stopping profiles were simulated using
\texttt{TRIM.SP}~\cite{1991-Eckstein-SSMS-10,trimsp}
in \qty{500}{\electronvolt}
increments for implantation energies between \qtyrange{0.5}{30}{\kilo\electronvolt}.
Each profile was fit using \Cref{eq:stopping,eq:beta-pdf}
and
the resulting ``shape'' parameters
(i.e., $\alpha_{i}$, $\beta_{i}$, $z_{\mathrm{max},i}$, and $f_{i}$)
were tabulated as a function of $E$.
To generate \emph{arbitrary} $\rho(z,E)$s within
$\qty{0.5}{\kilo\electronvolt} \leq E \leq \qty{30}{\kilo\electronvolt}$,
simple (linear) interpolation of the shape parameters was used.
As is evident in \Cref{fig:implantation-profiles},
this procedure produces stopping profiles in excellent agreement with
the Monte Carlo simulations.
A tabulation of the shape parameters for \ch{Nb2O5}(\qty{5}{\nano\meter})/\ch{Nb}
is given in \Cref{tab:stopping-pars}.

Thus,
from \Cref{eq:stopping,eq:beta-pdf},
along with Equation~(4) in the main text,
it is straightforward to determine $B(z)$ from fitting,
along with the characteristic superconducting lengths
---
the London penetration depth $\lambda_{L}$~\cite{1935-London-PRSLA-149-71}
and 
Pippard/\gls{bcs} coherence length $\xi_{0}$~\cite{1953-Pippard-PRSLA-216-547,1957-Bardeen-PR-108-1175}.

In the case that local electrodynamics are assumed,
Equation~(4) in the main text reduces to the analytic expression:
\begin{equation}
	\label{eq:london}
	B(z) = \tilde{B}_{0} \times \begin{cases}
		1, & z < d , \\
		\displaystyle \exp \left \{ -\frac{ ( z - d ) }{ \lambda(T) } \right \} , & z \geq d , 
	\end{cases}
\end{equation} 
where
$\tilde{B}_{0}$ is the effective external field (see Equation~(4) in the main text),
$d$ is the thickness of a non-superconducting surface ``dead layer''
(i.e., where $\tilde{B}_{0}$ isn't screened),
and
$\lambda(T)$ is the (effective) magnetic penetration depth~\cite{1996-Tinkham-IS-2}:
\begin{equation}
	\label{eq:lambda-two-fluid}
	\lambda(T) \approx \frac{ \lambda_{0} }{ \sqrt{1 - \left ( T / T_{c} \right )^{4} } } ,
\end{equation}
where $\lambda_{0}$ is its value extrapolated to \qty{0}{\kelvin}
[cf.\ the main text's Eq.~(7)].
While $\lambda_{L}$ and $\xi_{0}$ are explicit parameters in the main text's Equation~(4)
[via Equations~(6), (7), and (8)],
to determine their values in the local limit
we use $\lambda_{0}$'s well-known dependence on
the electron mean-free-path $\ell$
given by Equation~(10) in the main text.
Note that the factor $\pi/2$ included in this expression is in accord with
\gls{bcs} theory~\cite{1957-Bardeen-PR-108-1175}
for impure superconductors~\cite{1959-Miller-PR-113-1209,1971-Halbritter-ZP-243-201}.

\subsection{
	\label{sec:methods:direct}
	``Direct'' Meissner Screening Analysis
}

\begin{figure*}[p]
\centering
\newcommand{\customwidth}{0.31\textwidth}
\includegraphics[width=\customwidth]{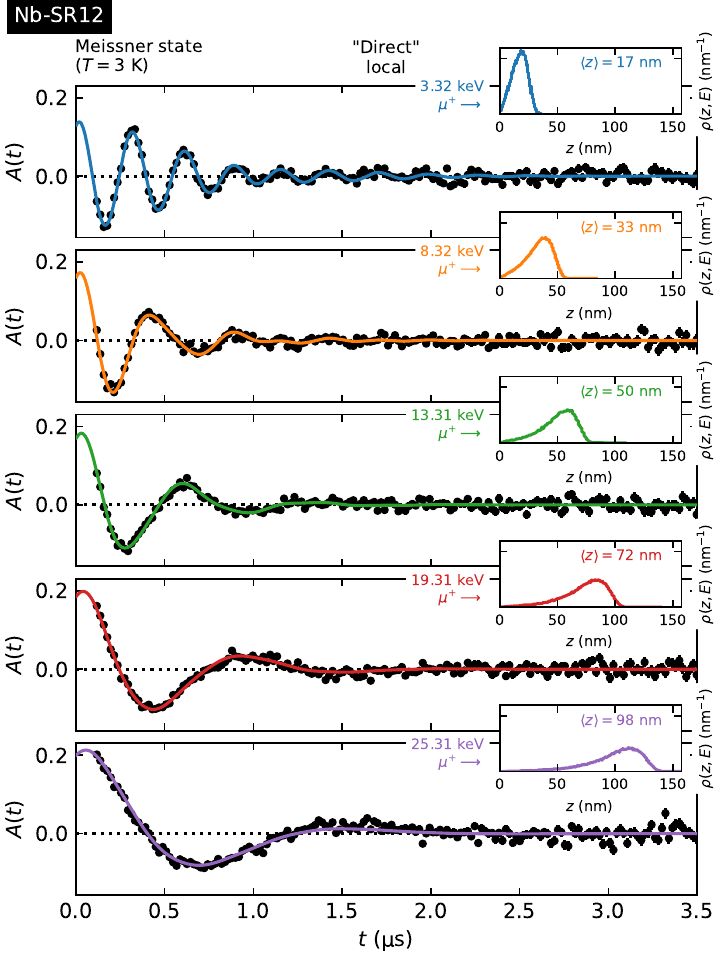}
\hfill
\includegraphics[width=\customwidth]{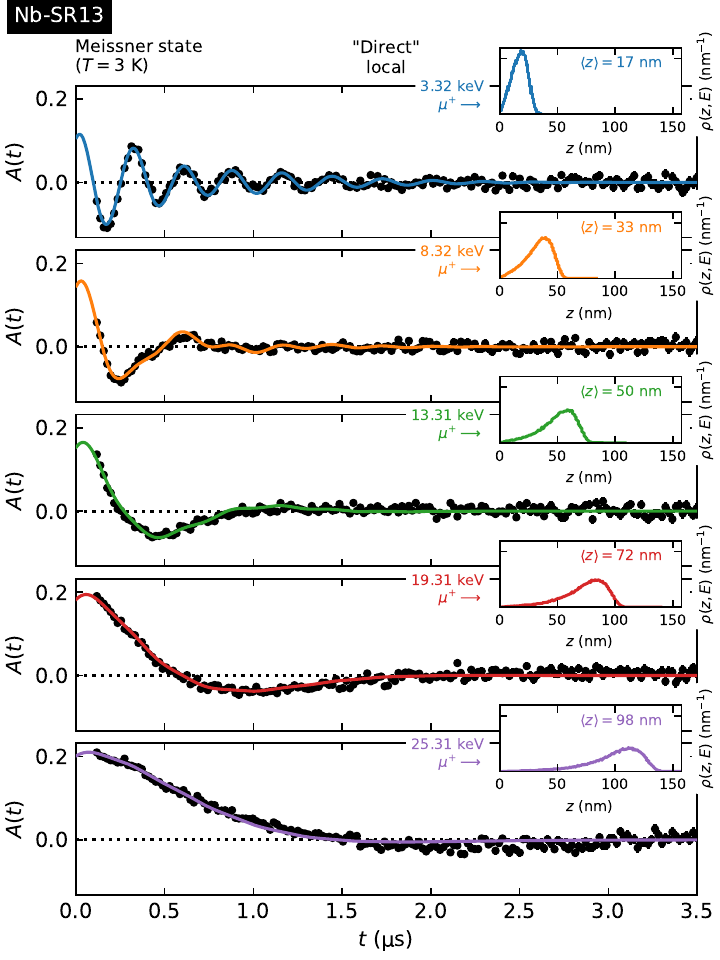}
\hfill
\includegraphics[width=\customwidth]{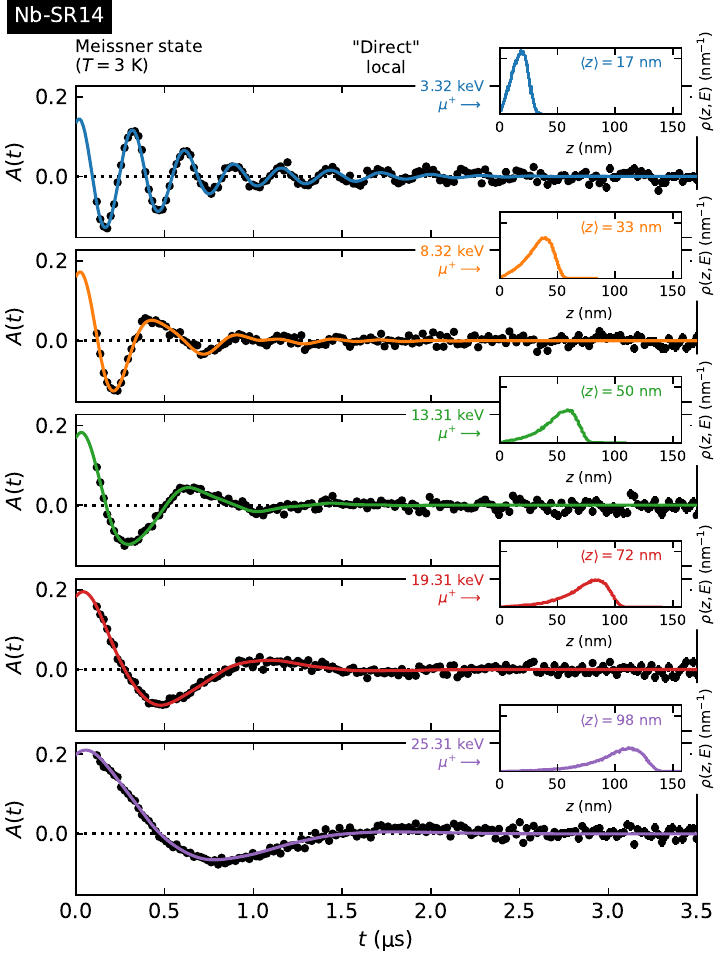} \\
\includegraphics[width=\customwidth]{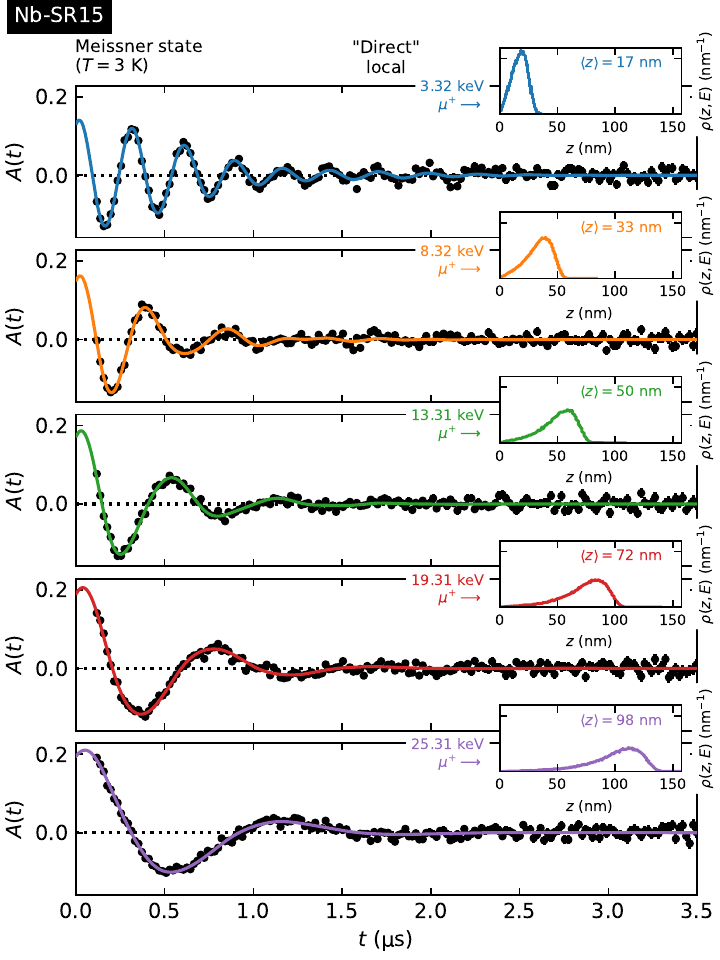}
\hfill
\includegraphics[width=\customwidth]{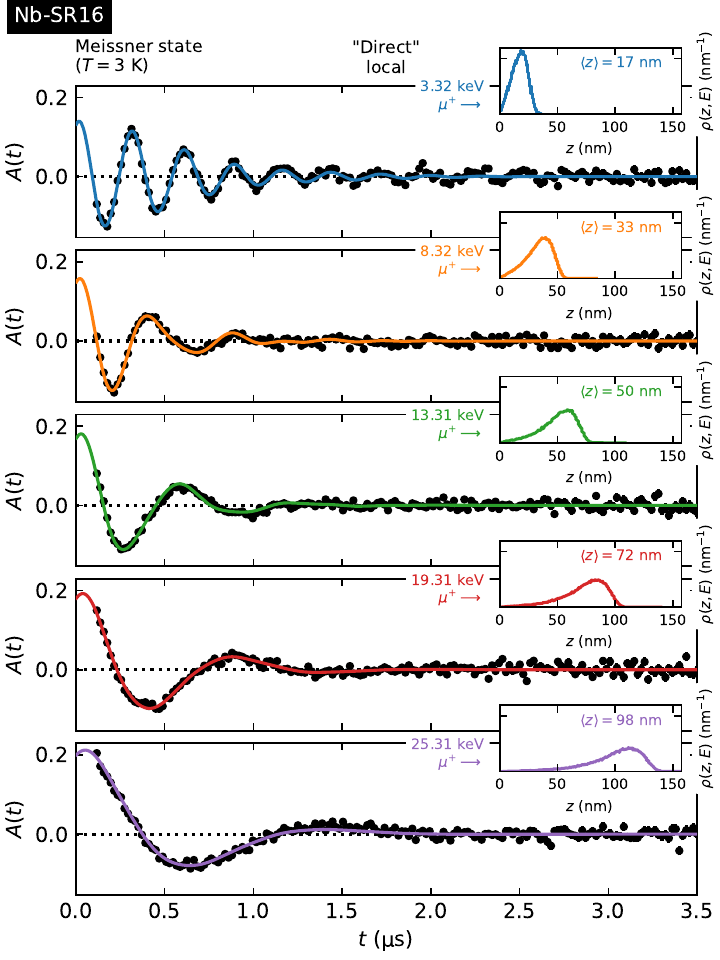}
\hfill
\includegraphics[width=\customwidth]{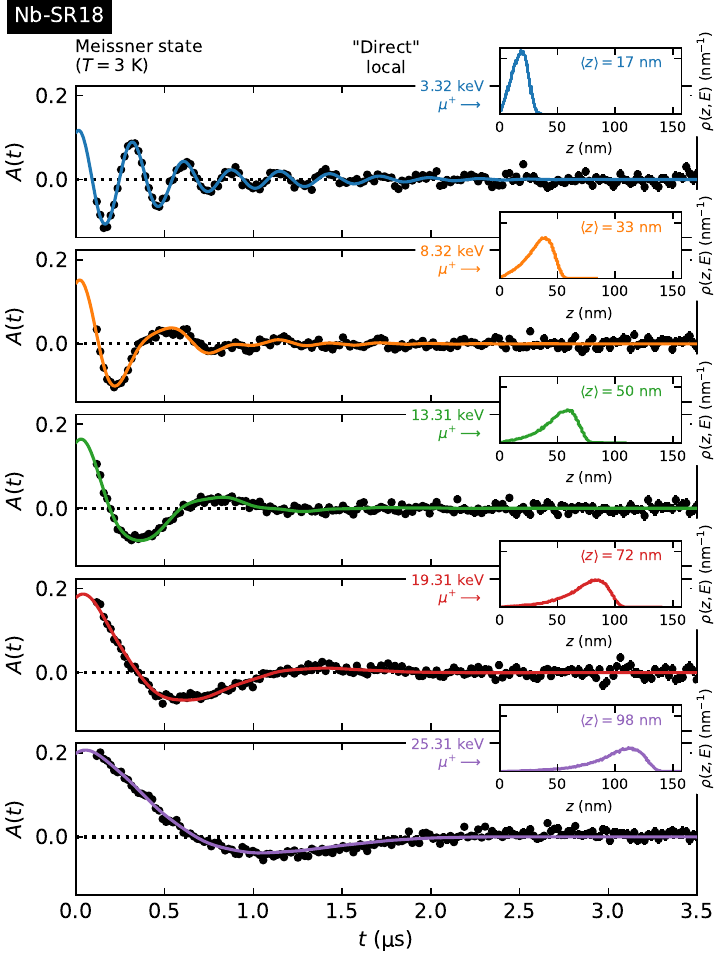}
\caption[
	\acrshort{le-musr} data in the \ch{Nb} samples fit using the ``direct'' approach
	and local electrodynamics.
]{
	\label{fig:asymmetry-direct-local}
	Typical \gls{le-musr} data in all \ch{Nb} samples fit using the ``direct'' approach
	and local electrodynamics.
	In the Meissner state,
	the damping of $A(t)$ is strong, increasing with increasing $E$,
	which is accompanied by a decrease in the
	rate of spin-precession.
	The solid colored lines denote fits using the model given by \Cref{eq:polarization-direct,eq:london},
	along with Equation~(4) in the main text.
	The $\mu^{+}$ implantation profile $\rho(z, E)$
	and
	mean stopping depth $\langle z \rangle$,
	simulated using the
	\texttt{TRIM.SP} Monte Carlo code~\cite{1991-Eckstein-SSMS-10,trimsp},
	are shown in the inset for each $E$.
}
\end{figure*}

\begin{figure*}[p]
\centering
\newcommand{\customwidth}{0.31\textwidth}
\includegraphics[width=\customwidth]{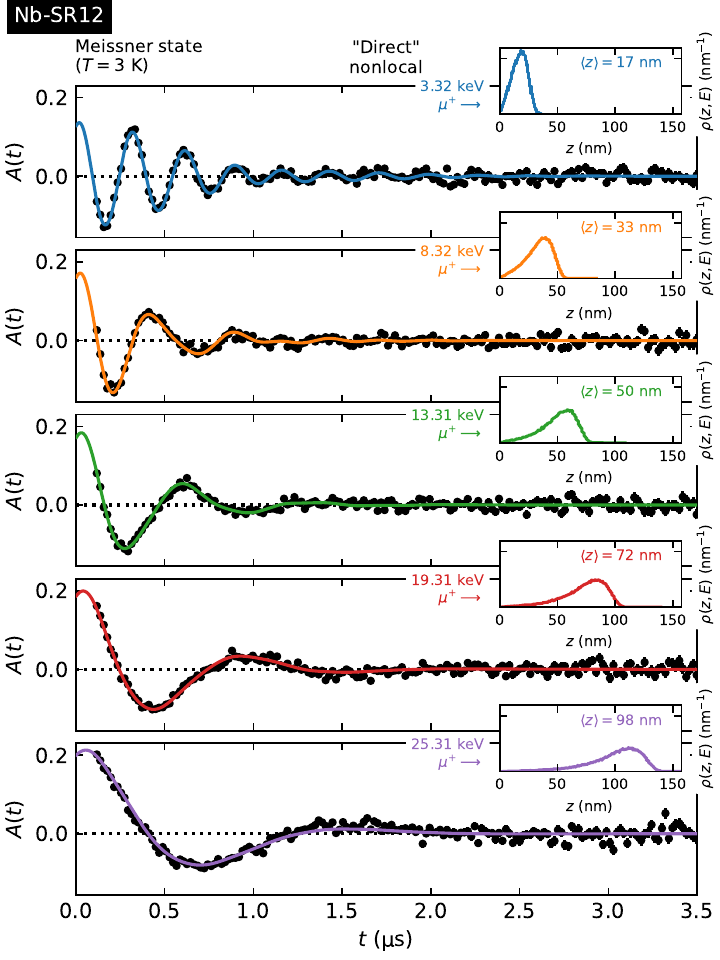}
\hfill
\includegraphics[width=\customwidth]{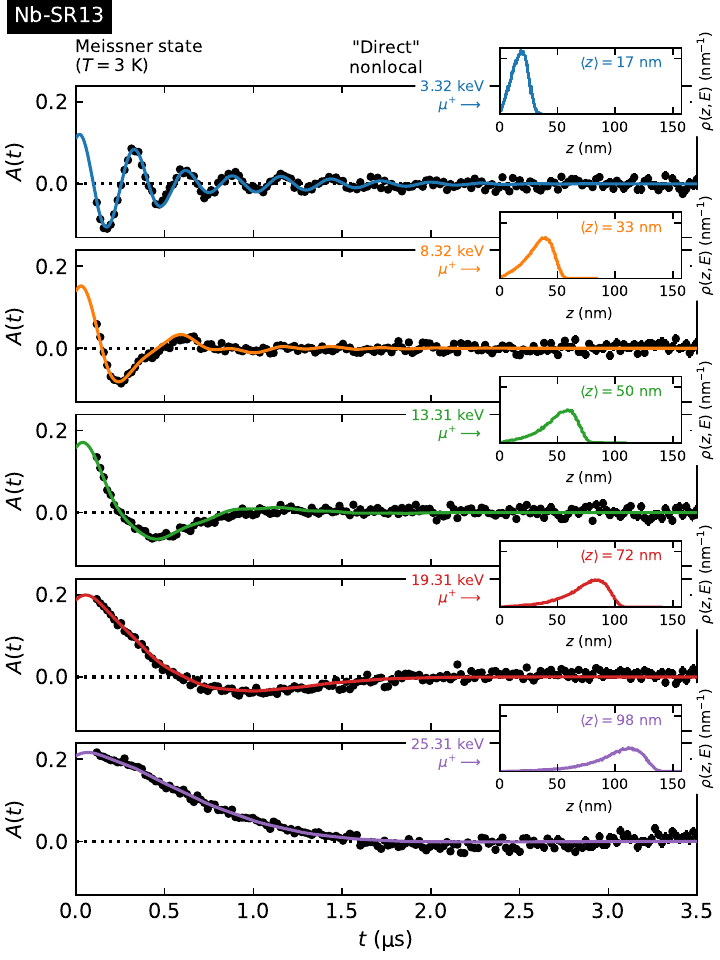}
\hfill
\includegraphics[width=\customwidth]{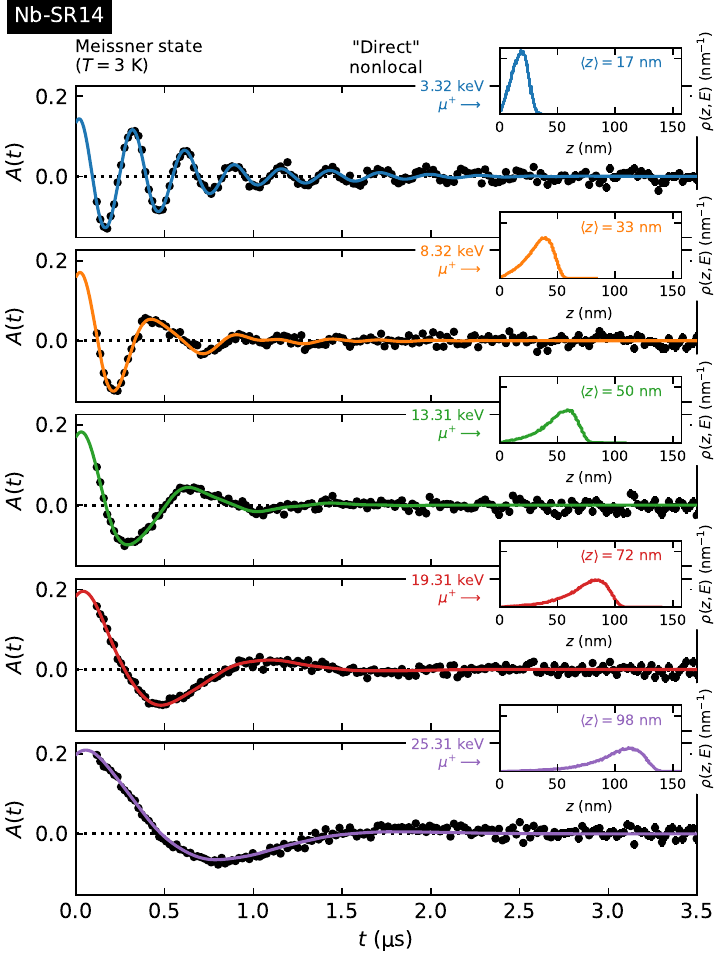} \\
\includegraphics[width=\customwidth]{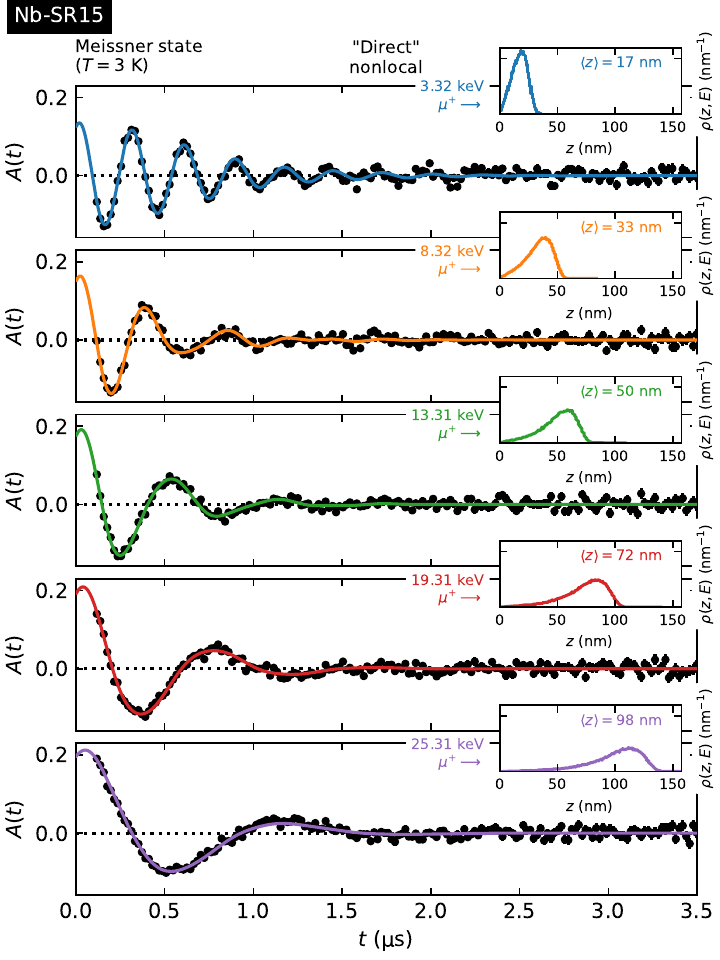}
\hfill
\includegraphics[width=\customwidth]{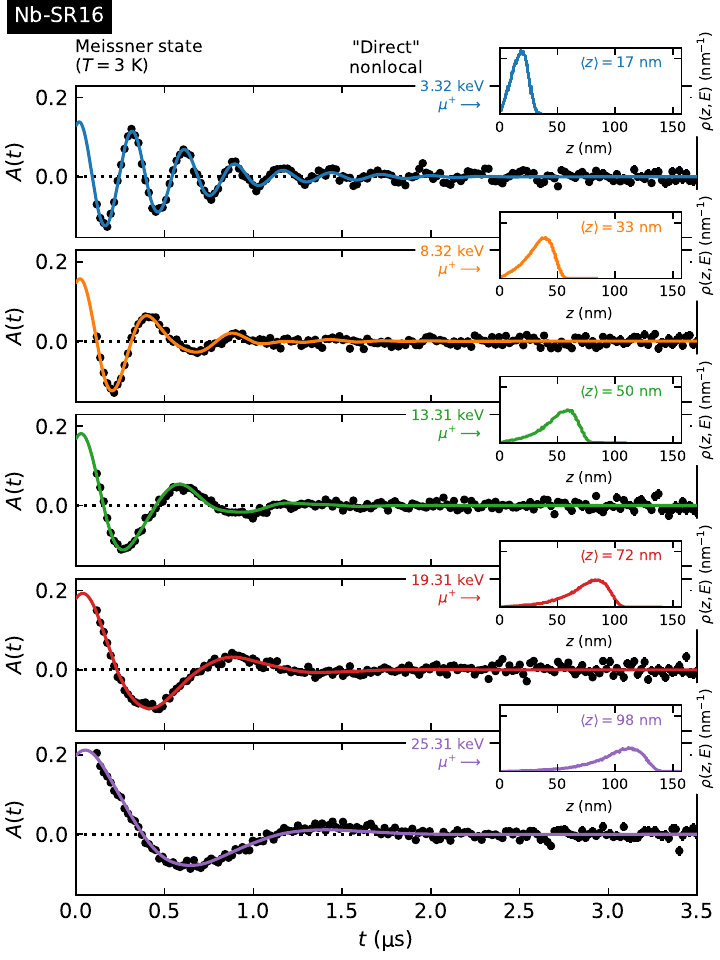}
\hfill
\includegraphics[width=\customwidth]{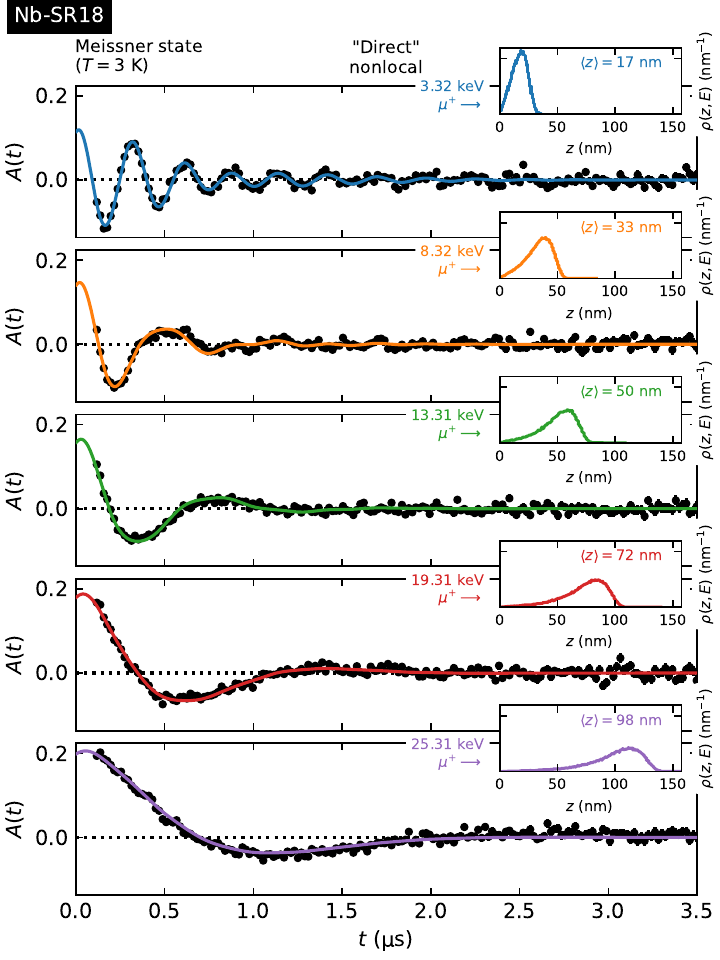}
\caption[
	\acrshort{le-musr} data in the \ch{Nb} samples fit using the ``direct'' approach
	and nonlocal electrodynamics.
]{
	\label{fig:asymmetry-direct-nonlocal}
	Typical \gls{le-musr} data in all \ch{Nb} samples fit using the ``direct'' approach
	and nonlocal electrodynamics.
	In the Meissner state,
	the damping of $A(t)$ is strong, increasing with increasing $E$,
	which is accompanied by a decrease in the
	rate of spin-precession.
	The solid colored lines denote fits using the model given by \Cref{eq:polarization-direct},
	along with Equations~(4) to (9) in the main text.
	The $\mu^{+}$ implantation profile $\rho(z, E)$
	and
	mean stopping depth $\langle z \rangle$,
	simulated using the
	\texttt{TRIM.SP} Monte Carlo code~\cite{1991-Eckstein-SSMS-10,trimsp},
	are shown in the inset for each $E$.
}
\end{figure*}

While the analysis described in \Cref{sec:methods:staged} and the main text
divides the extraction of the superconducting lengths into discrete ``stages,''
it is also possible to accomplishing this task \emph{directly} when fitting the raw \gls{le-musr} data.
As described elsewhere
(see, e.g., Refs.~\cite{2006-Suter-PB-374-243,2010-Kiefl-PRB-81-180502,2013-Kozhevnikov-PRB-87-104508}),
\Cref{eq:polarization} can also be calculated using:
\begin{widetext}
\begin{equation}
	\label{eq:polarization-direct}
	P_{\mu}(t) = \exp \left ( -\frac{\sigma_{b}^{2}t^{2}}{2} \right ) \int_{0}^{\infty} \rho(z, E) \cos \left ( \gamma_{\mu} B(z) t + \phi_{\pm} \right ) \, \mathrm{d}z ,
\end{equation}
\end{widetext}
where $B(z)$ is given by Equation~(3) or \Cref{eq:london}
(i.e., for the nonlocal and local limits, respectively),
and
$\sigma_{b}$ accounts for any ``extra'' damping caused by
inhomogeneous broadening~\cite{2010-Kiefl-PRB-81-180502}.
Fits following this approach can be performed using 
\texttt{musrfit}~\cite{2012-Suter-PP-30-69},
which uses the stopping profiles output by
\texttt{TRIM.SP}~\cite{1991-Eckstein-SSMS-10,trimsp}
as inputs when evaluating \Cref{eq:polarization-direct}.
A subset of the fit results are shown in
\Cref{fig:asymmetry-direct-local,fig:asymmetry-direct-nonlocal}
for the local and nonlocal limits, respectively.
Note that when local electrodynamics are assumed,
just as in the ``staged'' analysis,
$\lambda_{L}$ and $\xi_{0}$ are extracted
from $\lambda_{0}$'s dependence on $\ell$
[see Equation~(10) in the main text].

\section{
	Results
	\label{sec:results}
}

\subsection{
	Impurity Concentration \& Electron Mean-Free-Path
	\label{sec:results:impurity}
}

\begin{figure*}[p]
	\centering
	\newcommand{\customwidth}{0.31\textwidth}
	\includegraphics[width=\customwidth]{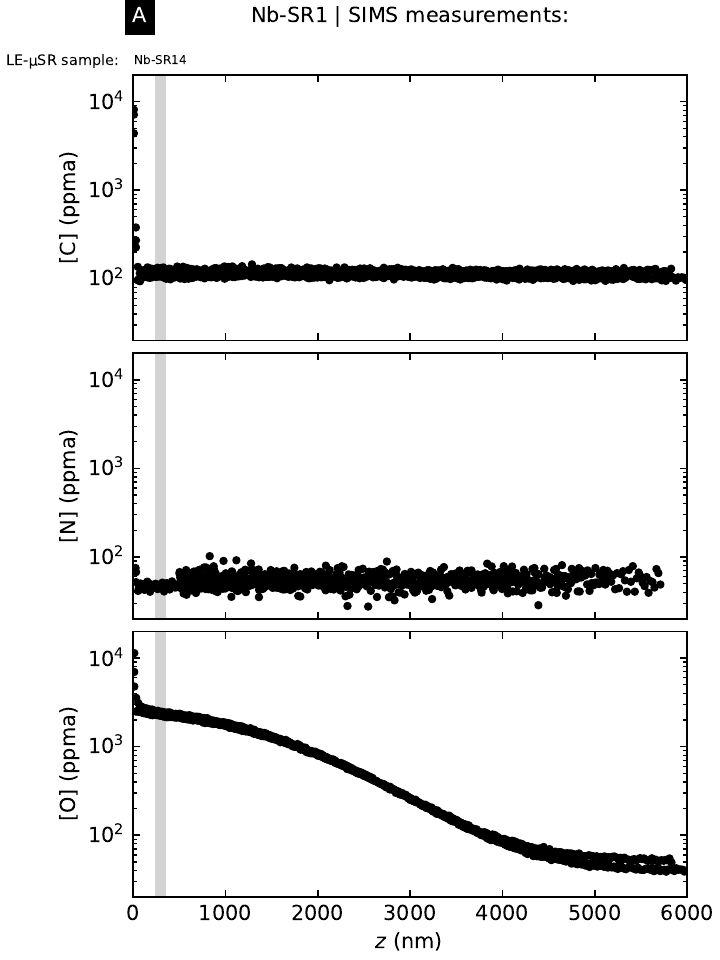} \hfill 
	\includegraphics[width=\customwidth]{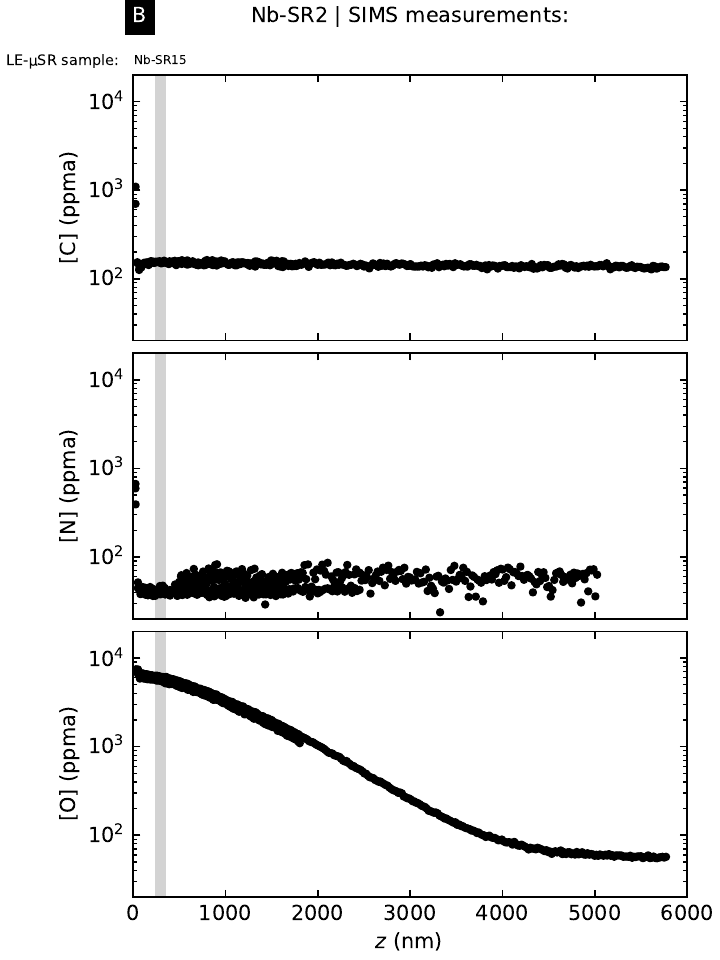} \hfill 
	\includegraphics[width=\customwidth]{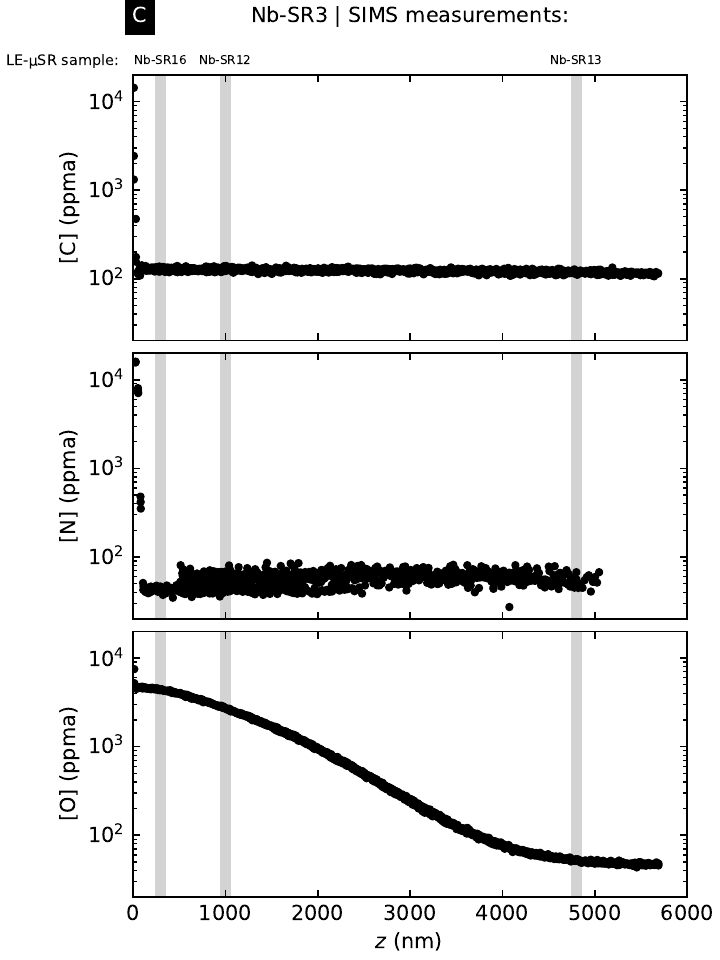} \hfill 
	\includegraphics[width=\customwidth]{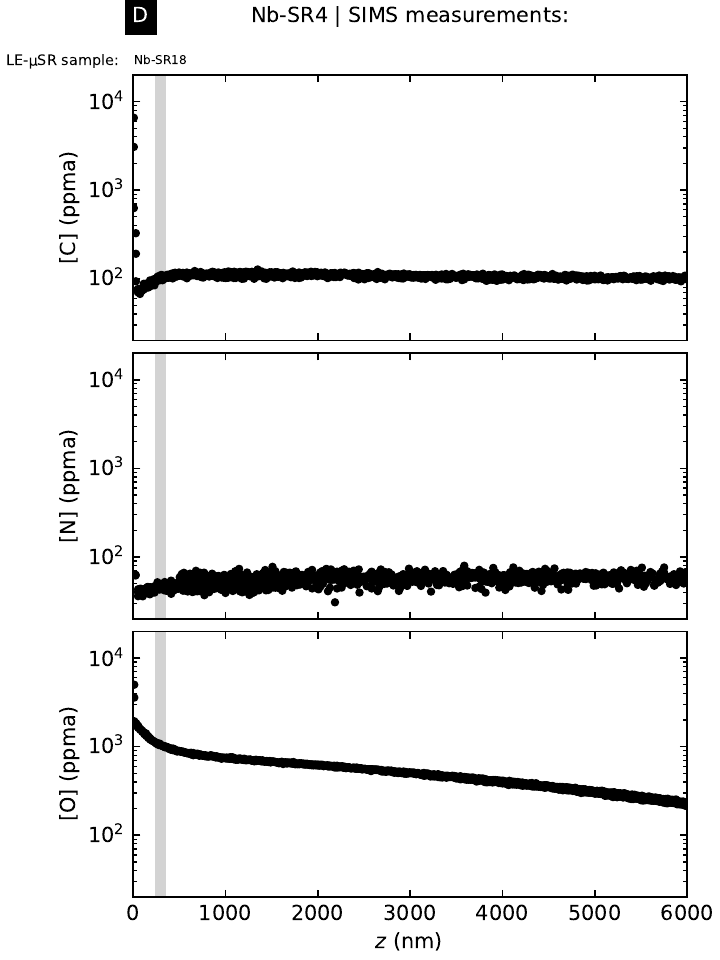} \qquad 
	\includegraphics[width=\customwidth]{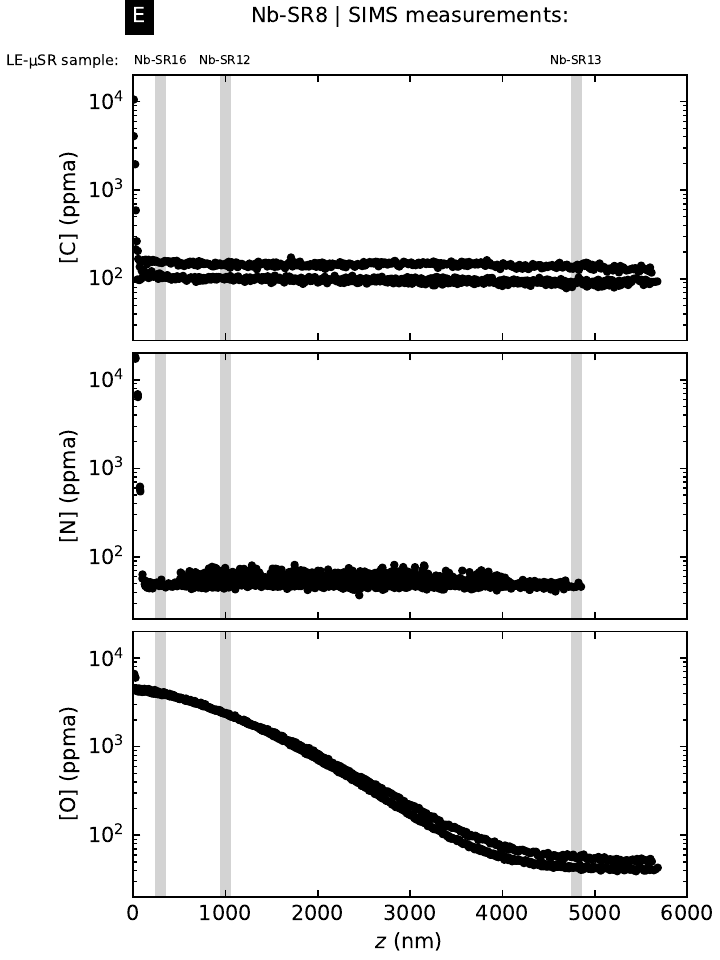}
	\caption[Impurity concentration profiles determined by \acrshort{sims} for the companion samples.]{
		\label{fig:impurities}
		Impurity concentration profiles determined by \gls{sims} for the companion samples:
		(\textbf{A}) Nb-SR1,
		(\textbf{B}) Nb-SR2,
		(\textbf{C}) Nb-SR3,
		(\textbf{D}) Nb-SR4,
		and
		(\textbf{E}) Nb-SR8.
		The grey shaded regions highlight \qty{100}{\nano\meter} windows
		at depths corresponding to the amount of surface material removed by
		\gls{ep} in the associated \gls{le-musr} samples
		(see \Cref{tab:samples}).
		Note that to estimate the impurity concent in sample Nb-SR13 sample,
		where no oxygen doping and additional surface \gls{ep} was performed,
		we use the ``baseline'' values of the profiles far below the surface
		(i.e., at depths $z \approx \qty{4.8}{\micro\meter}$) in \ch{Nb}-SR3/\ch{Nb}-SR8.
		A quantitative summary of the depth-dependent impurity values is given
		in \Cref{tab:sims-results-detailed}.
	}
\end{figure*}

Impurity profiles for \ch{C}, \ch{N}, and \ch{O} in several companion samples
(\ch{Nb}-SR1, \ch{Nb}-SR2, \ch{Nb}-SR3, \ch{Nb}-SR4, and \ch{Nb}-SR8)
determined from \gls{sims} measurements
are shown in \Cref{fig:impurities}.
Note that,
unlike the samples used in the \gls{le-musr} measurements,
none of the companion samples employed in the \gls{sims} measurements
received a final \gls{ep} treatment for surface removal
(see ``Step 8'' in \Cref{sec:methods:samples} and \Cref{tab:samples}).
Consequently,
impurity concentrations were estimated from the \gls{sims} profiles
over a \qty{\sim 100}{\nano\meter} window at depths corresponding to the
amount of surface material removed by \gls{ep}.
As the \gls{le-musr} sample Nb-SR13 did not receive any
surface/heat treatment following ``Step 5,''
its impurity concentrations were estimated from the ``baseline''
values of the impurity profiles far below the surface
(i.e., at depths $z \approx \qty{4.8}{\micro\meter}$) in \ch{Nb}-SR3/\ch{Nb}-SR8.
The statistical average of three profile measurement
``trials'' are used to estimate the impurity content within the \gls{ep} depth ``window.''
Due to the identical preparation of companion samples \ch{Nb}-SR3 and \ch{Nb}-SR8,
we also computed their combined average at each \gls{ep} depth ``window.''
A detailed summary of the extracted impurity values
is given in \Cref{tab:sims-results-detailed}.

\begin{table*}[p]
\centering
\caption[
	Summary of the impurity concentrations deduced from \acrshort{sims}
    in each companion sample.
]{
	\label{tab:sims-results-detailed}
	Summary of the impurity concentrations deduced from \acrshort{sims}
    in each companion sample at depths corresponding to the amount of surface material
    removed in the associated \gls{le-musr}.
    Concentrations representing a combined average comprising
	several measurement trials are given.
    As companion samples Nb-SR3 and Nb-SR8 both underwent identical surface treatments,
    we additionally provide impurity estimates from an average across each
    sample and its datasets.
    Corresponding values for the electron mean-free-path $\ell$,
	calculated using \Cref{eq:mfp-rrr,eq:residual-resistivity},
	are also given.
	Values used in the analysis of the \gls{le-musr} data appear in
	Table~I in the main text.
}
\begin{ruledtabular}
\begin{tabular}{l S l l S S S S}
{SIMS Sample} & {Depth (\unit{\nano\meter})} & {LE-$\mu$SR Sample} & {Trial} & {[\ch{C}] (\unit{ppma})} & {[\ch{N}] (\unit{ppma})} & {[\ch{O}] (\unit{ppma})} & {$\ell$ (\unit{\nano\meter})}  \\
\hline
Nb-SR1 & 300 & Nb-SR14 & \emph{Combined} & 115 \pm 11 & 45.4 \pm 2.9 & 2310 \pm 120 & 33.3 \pm 2.6 \\[1.5ex]
Nb-SR2 & 300 & Nb-SR15 & \emph{Combined} & 154.1 \pm 3.1 & 41.3 \pm 3.3 & 5840 \pm 350 & 13.6 \pm 1.2 \\[1.5ex]
Nb-SR3 & 300 & Nb-SR16 & \emph{Combined} & 127 \pm 6 & 43.0 \pm 2.8 & 4400 \pm 90 & 18.0 \pm 1.2 \\[1.5ex]
 & 1000 & Nb-SR12 & \emph{Combined} & 128 \pm 7 & 53 \pm 10 & 2720 \pm 90 & 28.3 \pm 2.0 \\[1.5ex]
 & 4800 & Nb-SR13 & \emph{Combined} & 118 \pm 4 & 53 \pm 7 & 51.8 \pm 1.1 & 365 \pm 24 \\[1.5ex]
Nb-SR4 & 300 & Nb-SR18 & \emph{Combined} & 101 \pm 6 & 46.2 \pm 3.2 & 1041 \pm 34 & 69 \pm 5 \\[1.5ex]
Nb-SR8 & 300 & Nb-SR16 & \emph{Combined} & 126 \pm 23 & 48.8 \pm 2.4 & 4040 \pm 120 & 19.5 \pm 1.4 \\[1.5ex]
 & 1000 & Nb-SR12 & \emph{Combined} & 119 \pm 22 & 55 \pm 8 & 2370 \pm 60 & 32.3 \pm 2.2 \\[1.5ex]
 & 4800 & Nb-SR13 & \emph{Combined} & 108 \pm 22 & 47.6 \pm 2.6 & 48 \pm 7 & 400 \pm 50 \\[1.5ex]
\hline \\[0.5ex]
Nb-SR3 + Nb-SR8 & 300 & Nb-SR16 & \emph{Combined} & 127 \pm 18 & 46 \pm 4 & 4200 \pm 210 & 18.8 \pm 1.5 \\[1.5ex]
 & 1000 & Nb-SR12 & \emph{Combined} & 123 \pm 17 & 54 \pm 9 & 2540 \pm 190 & 30.2 \pm 2.9 \\[1.5ex]
 & 4800 & Nb-SR13 & \emph{Combined} & 113 \pm 18 & 51 \pm 7 & 50 \pm 5 & 380 \pm 40 \\
\end{tabular}
\end{ruledtabular}

\end{table*}

As is well-known,
the electron mean-free-path
$\ell$ is directly related to the abundance of lattice defects
(e.g., impurity atoms, vacancies, dislocations, grain boundaries, etc.)
present in a metal.
In the limit of $T \rightarrow \qty{0}{\kelvin}$,
the most important contribution is generally impurity atoms,
which act as electronic scattering centers
and
raise \ch{Nb}'s residual resistivity.
This behavior can be expressed through the empirical relationship between
\ch{Nb}'s \gls{rrr} and $\ell$~\cite{1968-Goodman-JPF-29-240}:
\begin{equation}
	\label{eq:mfp-rrr}
	\ell = \frac{\sigma_{e}}{\rho_{e}} ,
\end{equation}
where $\rho_{e}$ is the sample's residual resistivity
and
$\sigma_{e} = \qty{3.7e-16}{\ohm\meter\squared}$~\cite{1968-Goodman-JPF-29-240}.
Noting that $\rho_{e}$ is the sum of contributions from all
impurity atoms,
it may be expressed as:
\begin{equation}
	\label{eq:residual-resistivity}
	\rho_{e} = \sum_{i} a_{i} \cdot [i] ,
\end{equation}
where $i$ denotes the impurity species
and $a_{i}$ denoting the (linear) proportionality with
concentration $[i]$.
Combining \Cref{eq:mfp-rrr,eq:residual-resistivity},
one obtains Equation~(1) in the main text.

While oxygen is the major ``contaminant'' in these experiments
(see \Cref{sec:methods:samples}),
we additionally account for the presence of other minor impurities
(i.e., \ch{C} and \ch{N}).
Following the findings in Ref.~\citenum{1976-Schulze-ZM-67-737},
we use
$a_{\ch{C}} = \qty{4.3 \pm 0.4 e-12}{\ohm\meter\per\ppma}$,
$a_{\ch{N}} = \qty{5.2 \pm 0.3 e-12}{\ohm\meter\per\ppma}$,
and
$a_{\ch{O}} = \qty{4.5 \pm 0.3 e-12}{\ohm\meter\per\ppma}$
for our estimation of $\ell$.
Values in each \gls{sims} companion sample at each \gls{ep}
depth ``window'' are tabulated in \Cref{tab:sims-results-detailed},
with the final values used to analyze the \gls{le-musr} data
summarized in Table~1 in the main text.

\subsection{
	Meissner Screening Profiles
	\label{sec:results:profile}
}

\subsubsection{
	``Staged'' Analysis
	\label{sec:results:profile:staged}
}

\begin{table*}[p]
	\caption[Summary of the ``staged'' Meissner screening profile analysis assuming local electrodynamics.]{
		\label{tab:results-local}
		Summary of the ``staged'' Meissner screening profile analysis assuming local electrodynamics
		(see Figure~2 in the main text).
		Here,
		$T$ is the absolute temperature of the \gls{le-musr} measurements,
		$B_{\mathrm{applied}}$ is the applied magnetic field,
		$\tilde{N}$ is the sample's (effective) demagnetization factor,
		$d$ is the thickness of the non-superconducting ``dead layer'' at the sample's surface,
		$\lambda(T)$ is the magnetic penetration depth at finite temperature,
		and
		$\lambda_{0}$ is the penetration depth's value extrapolated to
		\qty{0}{\kelvin} using \Cref{eq:lambda-two-fluid}.
		Also included are each sample's
		carrier mean-free-path $\ell$ determined from their impurity content
		using \Cref{eq:mfp-rrr,eq:residual-resistivity}~\cite{1976-Schulze-ZM-67-737,1968-Goodman-JPF-29-240}
		(see \Cref{sec:results:impurity}).
		For columns with only a single value, that parameter was treated as shared across all samples (i.e., in a global fit).
	}
	\begin{tabular*}{\textwidth}{l @{\extracolsep{\fill}} S S S S S S S}
\botrule
{Sample} & {$\ell$ (nm)} & {$T$ (\si{\kelvin})} & {$B_{\mathrm{applied}}$ (\si{\milli\tesla})} & {$\tilde{N}$} & {$d$ (\si{\nano\meter})} & {$\lambda(T)$ (\si{\nano\meter})} & {$\lambda_{0}$ (\si{\nano\meter})} \\
\hline
Nb-SR12 & 30.2 \pm 2.9 & 3.02 &  \dots  &  \dots  & 13.2 \pm 0.5 & 56.9 \pm 1.2 & 56.6 \pm 1.2 \\
Nb-SR13 & 380 \pm 40 & 2.99 &  \dots  &  \dots  & 12.6 \pm 0.4 & 35.4 \pm 0.7 & 35.2 \pm 0.7 \\
Nb-SR14 & 33.3 \pm 2.6 & 3.00 & 25.1586 \pm 0.0016 & 0.051 \pm 0.004 & 12.5 \pm 0.5 & 54.7 \pm 1.4 & 54.4 \pm 1.4 \\
Nb-SR15 & 13.6 \pm 1.2 & 3.02 &  \dots  &  \dots  & 12.6 \pm 0.6 & 68.9 \pm 1.6 & 68.5 \pm 1.6 \\
Nb-SR16 & 18.8 \pm 1.5 & 3.00 &  \dots  &  \dots  & 13.0 \pm 0.5 & 63.6 \pm 1.3 & 63.3 \pm 1.3 \\
Nb-SR18 & 69 \pm 5 & 3.04 &  \dots  &  \dots  & 14.3 \pm 0.4 & 43.5 \pm 1.1 & 43.3 \pm 1.1 \\

\botrule
\end{tabular*}
\end{table*}

\begin{table*}[p]
	\caption[Summary of the ``staged'' Meissner screening profile analysis assuming nonlocal electrodynamics.]{
		\label{tab:results-nonlocal}
		Summary of the ``staged'' Meissner screening profile analysis assuming nonlocal electrodynamics
		(see Figure~2 in the main text).
		Here,
		$T$ is the absolute temperature of the \gls{le-musr} measurements,
		$B_{\mathrm{applied}}$ is the applied magnetic field,
		$\tilde{N}$ is the sample's (effective) demagnetization factor,
		$d$ is the thickness of the non-superconducting ``dead layer'' at the sample's surface,
		$\lambda_{L}$ is the London penetration depth,
		and
		$\xi_{0}$ is the \gls{bcs} coherence length.
		Also included are each sample's
		carrier mean-free-path $\ell$ determined from their impurity content
		using \Cref{eq:mfp-rrr,eq:residual-resistivity}~\cite{1976-Schulze-ZM-67-737,1968-Goodman-JPF-29-240}
		(see \Cref{sec:results:impurity}).
		For columns with only a single value, that parameter was treated as shared across all samples (i.e., in a global fit).
	}
	\begin{tabular*}{\textwidth}{l @{\extracolsep{\fill}} S S S S S S S}
\botrule
{Sample} & {$\ell$ (nm)} & {$T$ (\si{\kelvin})} & {$B_{\mathrm{applied}}$ (\si{\milli\tesla})} & {$\tilde{N}$} & {$d$ (\si{\nano\meter})} & {$\lambda_{L}$ (\si{\nano\meter})} & {$\xi_{0}$ (\si{\nano\meter})} \\
\hline
Nb-SR12 & 30.2 \pm 2.9 & 3.02 &  \dots  &  \dots  & 12.4 \pm 0.6 &  \dots  &  \dots  \\
Nb-SR13 & 380 \pm 40 & 2.99 &  \dots  &  \dots  & 10.7 \pm 0.4 &  \dots  &  \dots  \\
Nb-SR14 & 33.3 \pm 2.6 & 3.00 & 25.1586 \pm 0.0016 & 0.059 \pm 0.005 & 11.6 \pm 0.6 & 29.0 \pm 1.2 & 45 \pm 5 \\
Nb-SR15 & 13.6 \pm 1.2 & 3.02 &  \dots  &  \dots  & 11.1 \pm 0.7 &  \dots  &  \dots  \\
Nb-SR16 & 18.8 \pm 1.5 & 3.00 &  \dots  &  \dots  & 11.9 \pm 0.6 &  \dots  &  \dots  \\
Nb-SR18 & 69 \pm 5 & 3.04 &  \dots  &  \dots  & 12.9 \pm 0.5 &  \dots  &  \dots  \\

\botrule
\end{tabular*}
\end{table*}

A summary of sample details, measurement conditions,
and fit parameters determined from the ``staged'' analysis
are given in
\Cref{tab:results-local} (local electrodynamics)
and
\Cref{tab:results-nonlocal} (nonlocal electrodynamics).
As expected from the screening curves shown in the main text's Figure~2,
the consistency between the two limits is evident,
with good agreement across all samples.
As mentioned in \Cref{sec:methods:staged},
in the local limit $\lambda_{L}$ and $\xi_{0}$ are determined from
from $\lambda_{0}$'s dependence on $\ell$ using
Equation~(10) in the main text.
This relationship is shown explicitly in
the main text's Figure~3,
with corresponding superconducting length scales given in both the figure's inset
and Table~II.

\subsubsection{
	``Direct'' Analysis
	\label{sec:results:profile:direct}
}

\begin{figure}[p]
	\centering
	\includegraphics[width=1.0\columnwidth]{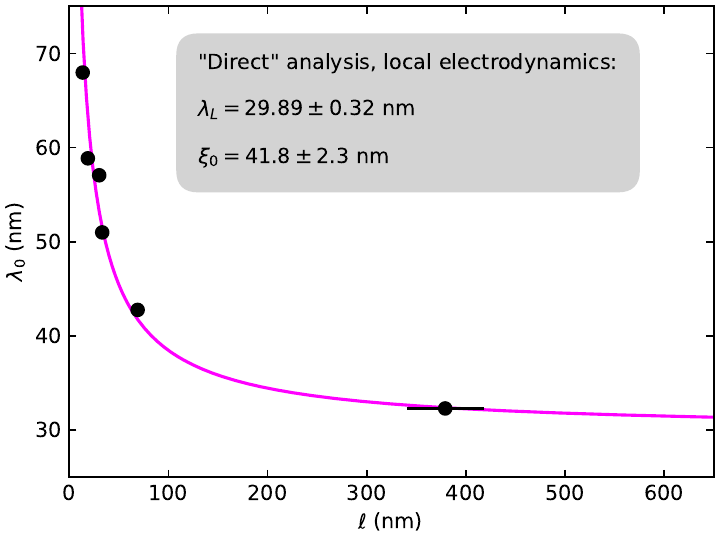}
	\caption[Determination of $\lambda_{L}$ and $\xi_{0}$ for the ``direct'' analysis approach in the local limit.]{
		\label{fig:lambda-vs-mfp-direct}
		Determination of the London penetration depth $\lambda_{L}$
		and
		Pippard/\gls{bcs} coherence length $\xi_{0}$
		for the ``direct'' analysis approach in the local limit.
		Shown is the relationship between the (effective)
		magnetic penetration depth at \qty{0}{\kelvin} $\lambda_{0}$
		and the carrier mean-free-path $\ell$.
		The solid coloured line denotes a best fit to Equation~(10) in the main text,
		with corresponding values for $\lambda_{L}$ and $\xi_{0}$
		indicated in the plot.
	}
\end{figure}

Results from the ``direct'' analysis,
described in \Cref{sec:methods:direct},
are summarized in
\Cref{tab:results-direct-local} (nonlocal electrodynamics)
and
\Cref{tab:results-direct-nonlocal} (nonlocal electrodynamics).
The $\lambda_{L}$ and $\xi_{0}$ values determined from this
approach compare well with those from the ``staged'' analysis
in \Cref{sec:results:profile:staged}.
As mentioned in \Cref{sec:methods:staged},
in the local limit $\lambda_{L}$ and $\xi_{0}$ are determined from
from $\lambda_{0}$'s dependence on $\ell$ using
the main text's Equation~(10).
This relationship is shown explicitly in
\Cref{fig:lambda-vs-mfp-direct},
with corresponding superconducting length scales given in the figure's inset,
along with Table~2 in the main text.

\begin{table*}[p]
	\caption[Summary of the ``direct'' Meissner screening profile analysis assuming local electrodynamics.]{
		\label{tab:results-direct-local}
		Summary of the ``direct'' Meissner screening profile analysis assuming local electrodynamics.
		Here,
		$T$ is the absolute temperature of the \gls{le-musr} measurements,
		$\tilde{B}_{0}$ is the effective magnetic field,
		$d$ is the thickness of the non-superconducting ``dead layer'' at the sample's surface,
		$\lambda(T)$ is the magnetic penetration depth at finite temperature,
		and
		$\lambda_{0}$ is the penetration depth's value extrapolated to
		\qty{0}{\kelvin} using \Cref{eq:lambda-two-fluid}.
		Also included are each sample's
		carrier mean-free-path $\ell$ determined from their impurity content
		using \Cref{eq:mfp-rrr,eq:residual-resistivity}~\cite{1976-Schulze-ZM-67-737,1968-Goodman-JPF-29-240}
		(see \Cref{sec:results:impurity}).
		The $\tilde{B}_{0}$s listed are equivalent to an effective demagnetization factor $\tilde{N}$ between \numrange{0.03}{0.05}
		(average value of \num{\sim 0.037}).
	}
	\begin{tabular*}{\textwidth}{l @{\extracolsep{\fill}} S S S S S S}
\botrule
{Sample} & {$\ell$ (nm)} & {$T$ (\si{\kelvin})} & {$\tilde{B}_{0}$ (\si{\milli\tesla})} & {$d$ (\si{\nano\meter})} & {$\lambda(T)$ (\si{\nano\meter})} & {$\lambda_{0}$ (\si{\nano\meter})} \\
\hline
Nb-SR12 & 30.2 \pm 2.9 & 3.02 & 262.8 \pm 0.4 & 13.71 \pm 0.14 & 57.41 \pm 0.12 & 57.08 \pm 0.12 \\
Nb-SR13 & 380 \pm 40 & 3.00 & 259.4 \pm 0.4 & 16.28 \pm 0.12 & 32.73 \pm 0.07 & 32.55 \pm 0.07 \\
Nb-SR14 & 33.3 \pm 2.6 & 3.00 & 260.70 \pm 0.26 & 14.82 \pm 0.14 & 51.47 \pm 0.11 & 51.18 \pm 0.11 \\
Nb-SR15 & 13.6 \pm 1.2 & 3.02 & 263.0 \pm 0.4 & 13.66 \pm 0.17 & 68.40 \pm 0.21 & 68.01 \pm 0.21 \\
Nb-SR16 & 18.8 \pm 1.5 & 3.00 & 260.78 \pm 0.26 & 14.51 \pm 0.13 & 59.68 \pm 0.20 & 59.35 \pm 0.20 \\
Nb-SR18 & 69 \pm 5 & 3.04 & 261.3 \pm 0.4 & 14.66 \pm 0.14 & 43.02 \pm 0.11 & 42.77 \pm 0.12 \\

\botrule
\end{tabular*}
\end{table*}

\begin{table*}[p]
	\caption[Summary of the ``direct'' Meissner screening profile analysis assuming nonlocal electrodynamics.]{
		\label{tab:results-direct-nonlocal}
		Summary of the ``direct'' Meissner screening profile analysis assuming nonlocal electrodynamics.
		Here,
		$T$ is the absolute temperature of the \gls{le-musr} measurements,
		$\tilde{B}_{0}$ is the effective magnetic field,
		$d$ is the thickness of the non-superconducting ``dead layer'' at the sample's surface,
		$\lambda_{L}$ is the London penetration depth,
		and
		$\xi_{0}$ is the Pippard/\gls{bcs} coherence length.
		Also included are each sample's
		carrier mean-free-path $\ell$ determined from their impurity content
		using \Cref{eq:mfp-rrr,eq:residual-resistivity}~\cite{1976-Schulze-ZM-67-737,1968-Goodman-JPF-29-240}
		(see \Cref{sec:results:impurity}).
		As the (statistical) error bars on both $\lambda_{L}$ and $\xi_{0}$ likely underestimate each
		parameter's true uncertainty,
		we report their (unweighted) averages in the table's bottom row,
		with an uncertainty given by their standard deviation.
		The $\tilde{B}_{0}$s listed are equivalent to an effective demagnetization factor $\tilde{N}$ between \numrange{0.03}{0.05}.
		(average value of \num{\sim 0.039}).
	}
	\begin{tabular*}{\textwidth}{l @{\extracolsep{\fill}} S S S S S S}
\botrule
{Sample} & {$\ell$ (nm)} & {$T$ (\si{\kelvin})} & {$\tilde{B}_{0}$ (\si{\milli\tesla})} & {$d$ (\si{\nano\meter})} & {$\lambda_{L}$ (\si{\nano\meter})} & {$\xi_{0}$ (\si{\nano\meter})} \\
\hline
Nb-SR12 & 30.2 \pm 2.9 & 3.02 & 262.73 \pm 0.21 & 13.0719 \pm 0.0024 & 31.92 \pm 0.05 & 39.7243 \pm 0.0028 \\
Nb-SR13 & 380 \pm 40 & 2.99 & 261.05 \pm 0.04 & 13.00 \pm 0.17 & 28.508 \pm 0.008 & 39.98 \pm 0.26 \\
Nb-SR14 & 33.3 \pm 2.6 & 3.00 & 261.64 \pm 0.18 & 13.68650 \pm 0.00024 & 29.86 \pm 0.04 & 38.7738 \pm 0.0008 \\
Nb-SR15 & 13.6 \pm 1.2 & 3.02 & 260.15 \pm 0.21 & 15.00 \pm 0.06 & 28.34 \pm 0.05 & 38.9254 \pm 0.0018 \\
Nb-SR16 & 18.8 \pm 1.5 & 3.00 & 260.75 \pm 0.18 & 14.5008 \pm 0.0022 & 26.23 \pm 0.04 & 32.275 \pm 0.004 \\
Nb-SR18 & 69 \pm 5 & 3.04 & 264.17 \pm 0.08 & 12.49 \pm 0.23 & 30.247 \pm 0.019 & 39.5 \pm 0.6 \\
\hline
 & & & & & 29.2 \pm 2.0 & 38.2 \pm 2.9 \\
\botrule
\end{tabular*}
\end{table*}

\bibliography{references-sm.bib}